\documentclass{aa}
\nolinenumbers
\usepackage[utf8]{inputenc}
\usepackage[varg]{txfonts}
\usepackage{amsmath}
\usepackage{xcolor}
\usepackage{multirow}
\usepackage{xspace}
\usepackage{xstring}
\usepackage{supertabular}
\usepackage{tabularx} 
\usepackage{orcidlink} 
\usepackage{placeins}
\usepackage{hyperref}
\hypersetup{colorlinks=true, linkcolor=blue, filecolor=magenta, urlcolor=cyan, citecolor=blue}

\title{Modelling non-radially propagating coronal mass ejections and forecasting the time of their arrival at Earth}
\titlerunning{Non-radially propagating halo-CMEs}
\author{A. Valentino\inst{\ref{cmpa}\,\orcidlink{0000-0001-9702-468X}}
\and J. Magdalenic\inst{\ref{cmpa},{\ref{rob}}\,\orcidlink{0000-0003-1169-3722}}}
\institute{Center for Mathematical Plasma Astrophysics, KU Leuven, Leuven 3000, Belgium \label{cmpa}
\and Royal Observatory of Belgium, 1180 Ukkel, Brussels, Belgium \label{rob}}

\abstract{We present the study of two solar eruptive events observed on December 7 2020 and October 28 2021. Both events were associated with full halo coronal mass ejections (CMEs) and flares. These events were chosen because they show a strong non-radial direction of propagation in the low corona and their main propagation direction observed in the inner heliosphere is not fully aligned with the Sun-Earth line. This characteristic makes them suitable for our study, which aims to inspect how the non-radial direction of propagation in the low corona affects the time of CMEs' arrival at Earth. We reconstructed the CMEs using SOHO/LASCO and STEREO/COR observations and modelled them with the 3D MHD model EUHFORIA and the cone model for CMEs. In order to compare the accuracy of forecasting the CME and the CME-driven shock arrival time at Earth obtained from different methods, we also used so-called type~II bursts, radio signatures of associated shocks, to find the velocities of the CME-driven shocks and forecast the time of their arrival at Earth. Additionally, we estimated the CME arrival time using the 2D CME velocity obtained from the white light images. 

Our results show that the lowest accuracy of estimated CME Earth arrival times is found when the 2D CME velocity is used (time difference between observed and modelled arrival time, $\Delta$t~$\approx$~-29~h and -39~h, for the two studied events, respectively). The velocity of the type~II radio bursts provides somewhat better — but still not very accurate — results ($\Delta$t~$\approx$~+21~h and -29~h, for the two studied events, respectively). Employing, as an input to EUHFORIA, the CME parameters obtained from the graduated cylindrical shell (GCS) fittings at consequently increasing heights, results in a strongly improved accuracy of the modelled CME and shock arrival time; $\Delta$t changes from 20~h to 10~min in the case of the first event, and from 12~h to 30~min in the case of the second one. 
This improvement shows that when we increased the heights of the GCS reconstruction we accounted for the change in the propagation direction of the studied CMEs, which allowed us to accurately model the CME flank encounter at Earth. 
Our results show the great importance of the change in the direction of propagation of the CME in the low corona when modelling CMEs and estimating the time of their arrival at Earth.} 


\begin{document}

\maketitle

\nolinenumbers

\section{Introduction} 
The most energetic phenomena that occur in the solar atmosphere are flares (e.g. \citealt{Shibata11}; \citealt[and references therein]{Benz17}) and coronal mass ejections (CMEs; see e.g. \citealt{Green18}; \citealt[and references therein]{Temmer21}). During flares, energy is released in the whole range of the electromagnetic spectrum (\citealt{Fletcher11}), while the main characteristic of CMEs is the ejection of magnetic structure into interplanetary space (e.g. \citealt{Cremades04}; \citealt{Temmer16}). Solar flares and CMEs are two processes that are generally closely related (\citealt{Vrsnak16}) in both time and space. In particular, the flare impulsive phase is often highly synchronous with the CME acceleration phase (see e.g. \citealt[and references therein]{Vrsnak08}). If the CMEs reach Earth they can cause geomagnetic storms. The severity of the storm depends on a number of different factors (\citealt{Vourlidas19}), with the most important being the CME's velocity, magnitude, and the orientation of the CME's magnetic field when it reaches Earth. In particular, different geomagnetic impact will be observed if the leading edge of the CME or the CME's flanks hit Earth's magnetosphere (\citealt{Koskinen06}; \citealt{Gopalswamy07}; \citealt{Scolini18}). So-called flank encounters, which usually have a weaker geomagnetic impact, are observed when the main propagation direction of the CME is not well aligned with the Sun-Earth line.

The CMEs can have velocities ranging from a few hundreds of km~s$^{-1}$ up to about 2\,500 km~s$^{-1}$ (\citealt{Vourlidas10}). The fast CMEs often drive the shock waves, observed as a faint structure in the white light (WL) images and also as a discontinuity in the solar wind parameters (\citealt{Reisenfeld03}; \citealt{Maloney11}; \citealt{Poedts16}). The indirect — but also the longest-known — signatures of the shock waves are so-called type~II radio bursts (\citealt{Wild50}; \citealt{Mann06}; \citealt{Vrsnak08}; \citealt{Magdalenic10}; \citealt{Magdalenic20}). The electrons accelerated at the shock front, driven by the CME or generated by the flare (e.g. \citealt{Magdalenic12}; \citealt{Magdalenic14}), can result in radio emission at the plasma frequency and/or its harmonic (\citealt{Magdalenic20}; \citealt{Jebaraj23c}). 

The change in the frequency over time — that is, the drift rate of the type~II bursts observed in dynamic radio spectra — can be  used as an indicator of the velocity of the shock and its driver. Type~II radio bursts mapping different heights, close to the Sun and in interplanetary space, are therefore often used to estimate the arrival of the shock wave and of the CME at Earth (\citealt{Cremades07}; \citealt{Cremades15b}; \citealt{Magdalenic08}; \citealt{Magdalenic12}; \citealt{Jebaraj20}). The type~II emission can originate from different parts of the same shock; that is, from the regions close to the CME's leading edge or from flank regions of the CME (see review by \citealt{Vrsnak08}; and also \citealt{Armatas19}; \citealt{Magdalenic20}; \citealt{Zucca18}; \citealt{Jebaraj20}; \citealt{Jebaraj21}). The relative position of the radio emission sources and the shock driver will therefore strongly influence the estimated shock arrival at Earth using the type~II bursts. The accuracy of the forecasted shock arrival at Earth will be particularly affected in the case of shocks associated with the CMEs that have a strongly non-radial direction of propagation in the low corona, and their main propagation direction in interplanetary space is not well aligned with the Sun-Earth line. Namely, if the radio emission originates from the CME flank that impacts Earth, we can expect that the forecasted arrival time using the radio bursts will be quite accurate. The forecasting accuracy is expected to decrease in the case of radio emission originating predominantly from the CME nose region or the second flank of the CME. One of the goals of our study is to test if this hypothesis can help us to understand the relative position of the shock wave and the type~II radio bursts. 

Our society is strongly dependent on satellites and ground infrastructure that can be impacted by the arrival of solar disturbances at Earth (see e.g. \citealt[and references therein]{Schrijver15}). Therefore, it has become very important to accurately predict the arrival time of CMEs and CME-driven shocks at Earth, and the effect that they would eventually have upon impact (\citealt{Webb00}; \citealt{Michalek06}). As a response to that need, over the last several decades a number of models have been developed with the aim of facilitating the forecasting of the CME and CME-driven shock wave arrival at Earth; for example, EUHFORIA (EUropean Heliospheric Forecast and Information Asset; \citealt{Pomoell18}, \citealt{Poedts20}), ENLIL (\citealt{Odstrcil99c}, \citealt{Odstrcil99b}), and SUSANOO (\citealt{Shiota16}). 
The main CME characteristics that are used as inputs to such models are mostly obtained in the low corona employing fitting procedures like, for example, the graduated cylindrical shell (GCS; \citealt{Thernisien06}; \citealt{Thernisien09}; \citealt{Thernisien11}), and often, but not always, assuming self-similar radial expansion from about 3-4~R$_\odot$ and up to 0.1~au. Such a treatment is based on studies that indicate that the larger part of the CME deflections and rotations happen at distances up to only a few solar radii (see e.g. \citealt{MacQueen86}; \citealt{Gosling87}; \citealt{Cremades04}; \citealt{Kay15}; \citealt{Mostl15}). 

In this study, we aim to understand how the assumed radial propagation direction of CMEs impacts the modelling accuracy of EUHFORIA. We consider two cases of strongly non-radially propagating CMEs in the low corona, and with the propagation direction in the inner heliosphere not well aligned with the Sun-Earth line. Additionally, we compare the accuracy of the modelling results with two other methods often employed in forecasting the CME and CME-driven shock arrival to Earth, such as the CME kinematics obtained from WL images and the aforementioned drift rates of type~II radio bursts.

We present a study of the CME and flare events on December 7 2020 and on October 28 2021. Both of these halo CMEs propagated towards Earth, although the extreme ultraviolet (EUV) and WL images showed that the CMEs have a strongly non-radial propagation direction in the low corona and that the bulk of the CME mass had a propagation direction which was not well aligned with the Sun-Earth line. 
The paper is structured as follows. In Sect.~\ref{Data}, we present the observations and the work methodology. 
A detailed description of events is provided in Sect.~\ref{Events}, which also contains all of the information about the modelling of the two studied events with EUHFORIA. 
Section~\ref{Results} discusses the results of the study presented and provides inputs for further improvements of the modelling of the timing of CMEs' arrival at Earth. 

\section{Observations and methods} \label{Data}

Our multi-wavelength study of two CME and flare events focuses on EUV, WL, radio, and in situ observations. These observations were complemented by modelling of the solar wind and the CMEs using the recently developed 3D magnetohydrodynamics (MHD) model EUHFORIA. In order to estimate the shock wave kinematics from radio observations, we employed different 1D density models. All employed observations and models are described in this section. 

\subsection{Observations} \label{Observations}

The flare time profiles were obtained from Geostationary Operational Environmental Satellite (GOES; \citealt{Garcia94}) observations. To understand the association of the flare and the CME and the dynamics of the CME's on-disc signatures (such as waves and dimmings) we also used EUV observations taken by the Atmospheric Imaging Assembly (AIA; \citealt{Lemen12}) on board the Solar Dynamics Observatory (SDO; \citealt{Pesnell12}). In particular, we employed SDO/AIA observations in three channels at 304\,\AA\,, corresponding to temperatures of the chromosphere and the transition region, and 193\,\AA\ and 211\,\AA\,, corresponding to the temperatures of the hot coronal plasma. Helioseismic and Magnetic Imager (HMI, also on board the SDO; \citealt{Scherrer12}) data were used to characterise the magnetic complexity of the active region (AR) from which the studied CME-flare events originated. These observations also helped us to determine the orientation of the magnetic loops at the time of eruptions, and the orientation of their neutral line on the photospheric level. The Global Oscillation Network Group (GONG) synoptic magnetogram maps \citep{Hill18} were used as the main input for modelling of the background solar wind with EUHFORIA. 

To study the evolution of CMEs in the inner corona, we employed coronagraphic data from the Large Angle and Spectroscopic COronagraph (LASCO; \citealt{Brueckner95}) instrument on board the SOlar and Heliospheric Observatory (SOHO; \citealt{Domingo95}). We mostly used observations of the SOHO/LASCO~C2 and C3 coronagraphs that provide observations in the field of view of 1.5-30~R$_\odot$ (solar radii). We also used WL observations from COR1 and COR2 coronagraphs \citep{Howard08} on board the Solar TErrestrial RElations Observatory Ahead (STEREO~A; \citealt{Kaiser08}), which have a combined field of view from 1.5-15~R$_\odot$. At the time of our events, the STEREO B spacecraft was not operational. The CME arrival at Earth was inspected in a high-resolution in situ time series (1~min) of the ambient plasma characteristics obtained from the \href{https://omniweb.gsfc.nasa.gov}{OMNI} webpage, which compiles the in situ observations from several spacecraft, such as the Advanced Composition Explorer (ACE, \citealt{Stone98}), WIND (\citealt{Harten95}), and the Deep Space Climate Observatory (DSCOVR). 

In order to verify that the observed discontinuities are indeed shocks, we employed the following criteria, as they are given in the \href{http://www.ipshocks.fi}{Helsinki Shock Waves} \citep{Kilpua15} database:

\begin{equation}
    V_{sh} = V_{down} - V_{up} \geq 20 km~s^{-1}, 
\end{equation} 

\begin{equation}
    N_{sh} = \frac{N_{p}^{down}}{N_{p}^{up}} \geq 1.2, 
\end{equation}

\begin{equation}
    T_{sh} = \frac{T_{p}^{down}}{T_{p}^{up}} \geq \frac{1}{1.2},
\end{equation} 

\begin{equation}
    B_{sh} = \frac{B_{down}}{B_{up}} \geq 1.2, 
\end{equation}

where \textit{V, N, T, B} define, respectively, the velocity, density, temperature, and magnetic field at the shock wave front. The subscript 'sh' stands for 'shock', and 'down' and 'up' stand for the downstream and upstream parts of the shock, respectively. 

In the study of the shock associated radio emission, we inspected all available ground- and space-based observations. Nevertheless, we focused on the space-based radio events as they cover similar radial distances to the coronagraph SOHO/LASCO~C2 observations. The space-based radio observations from the SWAVES instrument on board STEREO~A \citep{Bougeret08} and the WAVES experiment on board the Wind spacecraft \citep{Bougeret95} presenting two studied radio events are shown in Fig.~\ref{fig:swaves}. Both of those instruments provide dynamic radio spectra and direction-finding measurements at a number of discrete frequencies. The observations by STEREO/WAVES provide dynamic spectra in the range of 10\,-\,16\,000~kHz, and those by Wind/WAVES provide dynamic spectra in the range of 4\,-\,13\,825~kHz, by utilizing three different antennas at different ranges. 

\subsection{Methods} \label{methods}
In this study, we employed a few different methods and tools, with an emphasis on the two main methods of CME reconstruction and modelling: a) the GCS technique (\citealt{Thernisien06}, \citealt{Thernisien09}; \citealt{Thernisien11}), and b) EUHFORIA (\citealt{Pomoell18}; \citealt{Poedts20}), the inner heliospheric model of solar wind and CMEs. 

\begin{figure*}[htb]
    \centering
    \includegraphics[width=0.32\linewidth]{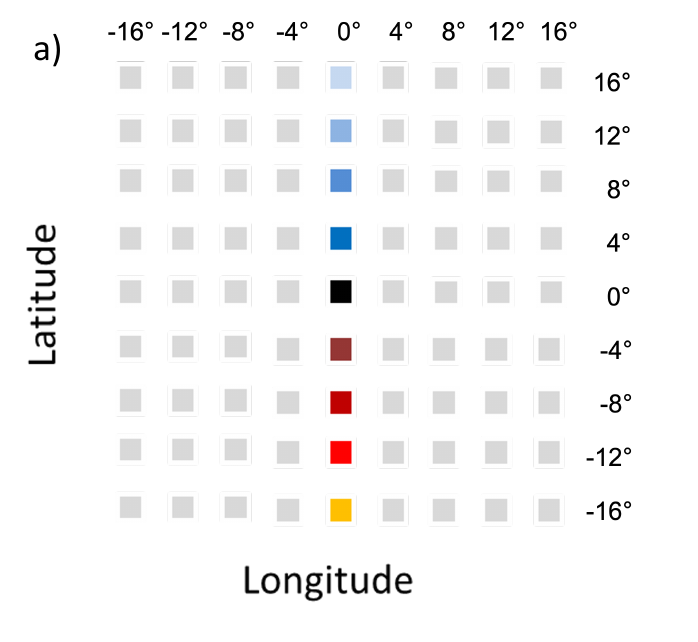} \includegraphics[width=0.60\linewidth]{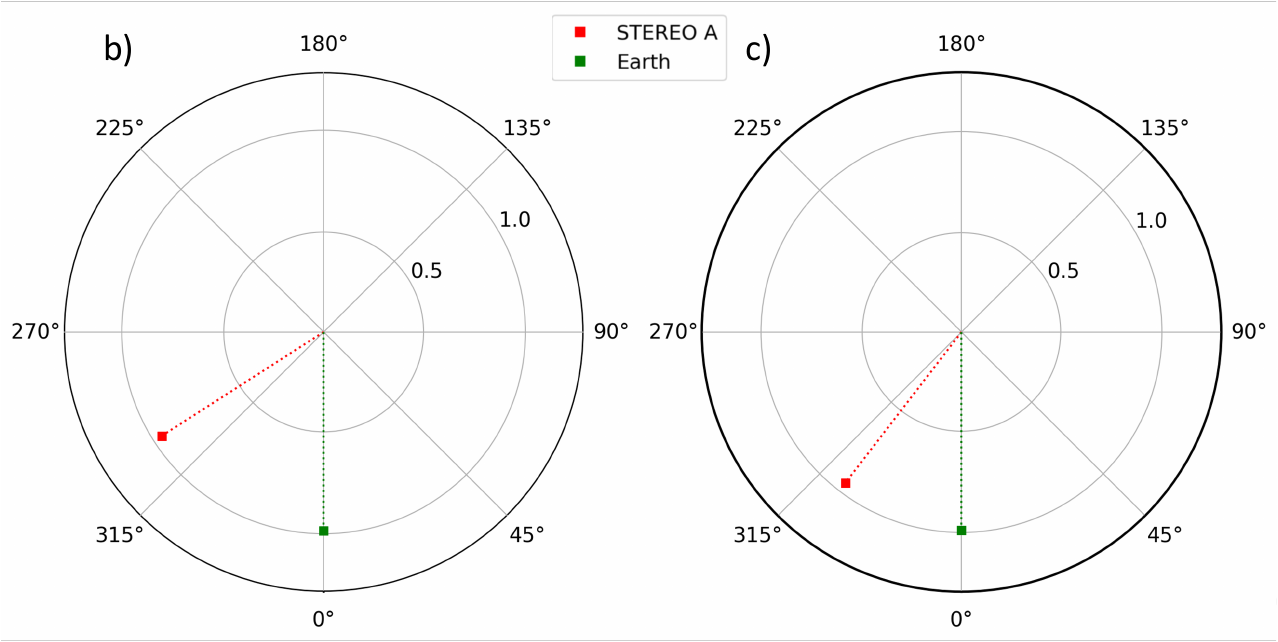}
    \caption{a) Grid of virtual spacecraft used for the EUHFORIA simulations. Each square represents a virtual spacecraft. The colored squares at zero longitude are those that we are interested in and those that are presented on the plots, with the respective colours. b) and c) Positions of the STEREO spacecraft with respect to Earth during Event~1 and Event~2, respectively.}
    \label{fig:STEREO_pos}
\end{figure*} 

The GCS method uses coronagraph images to fit a croissant-like 3D shell structure to the CME observed in WL in order to obtain the CME's geometric parameters. This shape consists of two cones, corresponding to the legs of the CME, with their narrowest part, the vertex, located at the center of the Sun, and a tubular section connecting the two cones and corresponding to the main body of the CME. This technique allows a number of parameters — for example, the longitude and latitude of the CME's source region — to be fitted in order to better match the croissant-like shell to the WL image. Applying this reconstruction technique to a series of images allows us to obtain information about the 3D CME kinematics. 
We note that a certain level of inaccuracy in the GCS fitting might arise from fitting the perfect croissant shape to the often diffuse CME structure observed in the WL. The fitting results are strongly dependent on the subjective view of the person performing the fitting (see e.g. \citealt{Verbeke23}).

In order to study the CME's propagation and estimate the time of its arrival at Earth, we used EUHFORIA, the 3D MHD model of the solar wind and of the CMEs\textbf{,} with the modelling domain up to distances of 2~au. EUHFORIA consists of three main parts: the coronal part, the heliospheric part, and the CME models. The inputs for the coronal part of the model are synoptic magnetograms. The most-often employed are GONG synoptic maps, but some other inputs such as ADAPT and HMI maps are also available (e.g. \citealt{Perri23}). The coronal model, which extends from the Solar surface up to the inner boundary of the heliospheric domain at 0.1~au, provides the initial parameters as an input to the heliospheric part of EUHFORIA at the inner boundary. 

The CMEs were also inserted at EUHFORIA's inner boundary. The modelling of CME propagation and evolution was done by using the input conditions of the heliospheric part and by solving the 3D time-dependent MHD equations, while also taking into consideration interactions with the background solar wind. The coordinate system in which those equations were solved is the Heliocentric Earth EQuatorial (HEEQ), in which the $Z$ axis is parallel to the rotational axis of the Sun and the $X$ axis connects the Sun and the Earth. 

For our simulations, we used the default set-up of EUHFORIA (see e.g. \citealt{Hinterreiter19}) and the cone model for the CMEs. The cone model considers a particular case of the croissant-like configuration of the CME and consists of two parts. A circular cone with its narrowest part, the vertex, located at the center of the Sun is combined with the top front part attached to the base of the cone. In this study, we used the cone-CME of a spheroidal shape, as it is defined in \citealt{Scolini18b}. The parameters needed for the description of the CME cone are the latitude and the longitude, which represent the point on the surface of the Sun that intersects with the line that connects the center of the leading edge of the CME with the vertex of the cone. Other important parameters are the half angular width and velocity of the CME. The cone model does not take into account the magnetic field, and therefore does not provide information about its evolution within the CME. It could be simply described as a blob of plasma that propagates in interplanetary space. This also means that the model does not provide any information about the passage of the magnetic cloud, following the shock, and we cannot distinguish in the in situ data between the CME-driven shock and the CME itself. To assess the arrival of the selected CMEs at Earth, we applied the forward-shock identification criteria used by the Heliospheric Shock Database developed and maintained at the University of Helsinki \citep{Kilpua15}. Such a description is sufficient for our study, which focuses on the time of the CME's arrival at Earth, and not on its geomagnetic impact upon arrival at Earth. The simplicity of the cone model also makes our study not too computationally demanding. As it is difficult to estimate the CME mass and density of individual events, we considered values generally employed in EUHFORIA: a density of $\rho_{CME} = 10^{-18}$~kg~cm$^{-3}$ and a temperature of T$_{CME}$=0.8~MK (\citealt{Pomoell18} and similar to \citealt{Scolini18}; \citealt{Scolini18b}; \citealt{Scolini20}; \citealt{Verbeke22}). The angular resolution used for EUHFORIA's coronal part is 2$^{\circ}$, while for the heliospheric part the radial resolution is 512 and the angular one is 2$^{\circ}$ per pixel. 

In order to be able to assess the arrival of the CME in the near-Earth environment, we positioned virtual spacecraft around Earth with a 4$^{\circ}$ spacing. Fig.~\ref{fig:STEREO_pos}a shows the grid of virtual spacecraft positioned in the modelling domain of EUHFORIA. Each square corresponds to a virtual spacecraft and they span between $\pm$~16$^{\circ}$, both in longitude and latitude with respect to Earth, which is at the center of the grid, at 0$^{\circ}$ longitude and 0$^{\circ}$ latitude, and which is marked with a black square. The virtual spacecraft considered in the study are primarily those at 0$^{\circ}$ longitude. The squares in shades of blue correspond to virtual spacecraft located above Earth and those in shades of red correspond to virtual spacecraft located below Earth. We used the same colouring to show the modelled time series in Figs.~\ref{fig:December_DONKI}, \ref{fig:GCS-fit4}, \ref{fig:Donki_Orig}, and \ref{fig:GCS-44}.

In order to have the most realistic conditions for the propagation of the modelled CMEs, we first optimised the solar wind. Different GONG magnetograms, at times prior to the eruption up to the maximum time window of 12 hours, were employed. The aim was to achieve the best agreement between the modelled and the observed in situ solar wind characteristics, to the point until just before the arrival of the CME-driven shock wave. The magnetogram that provided the best agreement between the modelled and the observed solar wind velocity was then used as an input to the heliospheric part of EUHFORIA.

Together with the advanced CME modelling, we also employed two simple methods of estimating the arrival time of the CME and CME-driven shock wave at Earth. The first method considers the drift rate of the type II radio bursts considered the drift rate of the type II radio burst signature of the shock waves, and the 1D coronal electron density models (\citealt{Newkirk61}; \citealt{Saito70}; \citealt{Leblanc98}; \citealt{Mann99}), in order to provide the velocity of the type~II radio bursts; in other words, the velocity of the associated shock wave. In this study, we employed the 1fold model by \cite{Leblanc98}, which is generally used for space-based observations. Details on such a way of estimating the shock wave velocity can be seen, for example, in \cite{Magdalenic08}, \cite{Magdalenic14}, \cite{Nindos08}, \cite{Nindos11}, \cite{Zucca14a}, and \cite{Zucca14b}. 

The second and most frequently applied method of estimating the time at which the CME arrives at Earth is based on the estimation of the CME and shock wave speed using the single view-point WL images. The height-time plots of the CME's leading edge are used to derive the velocity of the CME, whose values can also be obtained in the \href{https://cdaw.gsfc.nasa.gov/CME_list/index.html}{SOHO/LASCO} catalogue. In such a way, the obtained CME velocity can also be partially corrected for projection effects (\citealt{Reiner01}; \citealt{Reiner05}; \citealt{Schwenn05}; \citealt{Gopalswamy18}) and is considered to be a good first approximation for the estimation of the CME arrival time.

\begin{figure*}[htb]
    \centering
    \includegraphics[width=0.305\linewidth]{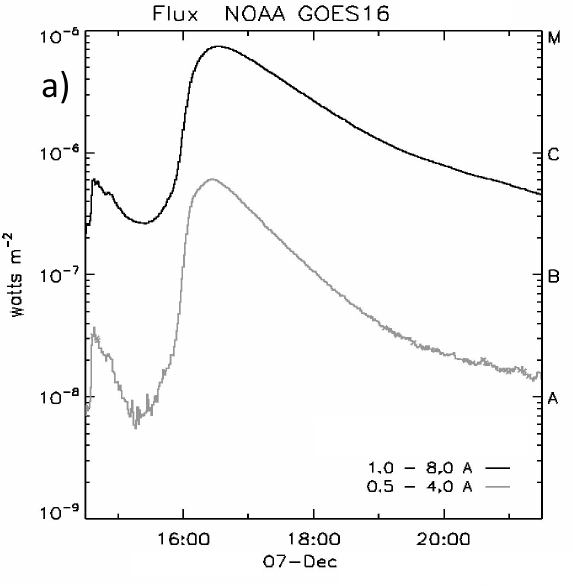}
    \includegraphics[width=0.31\linewidth]{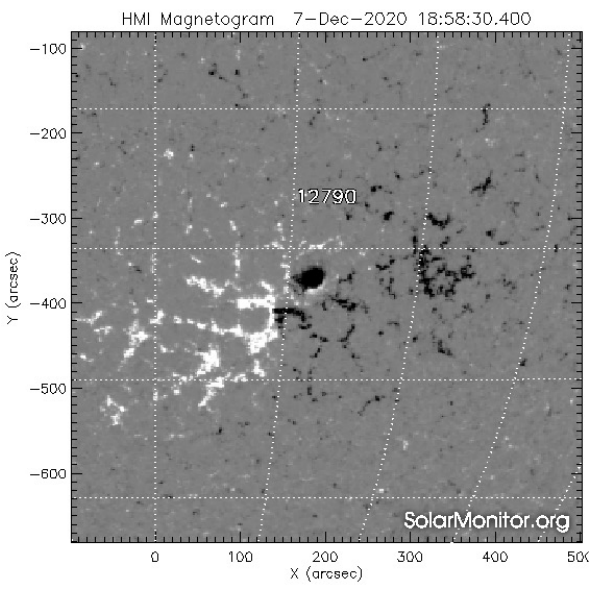}
    \includegraphics[width=0.32\linewidth]{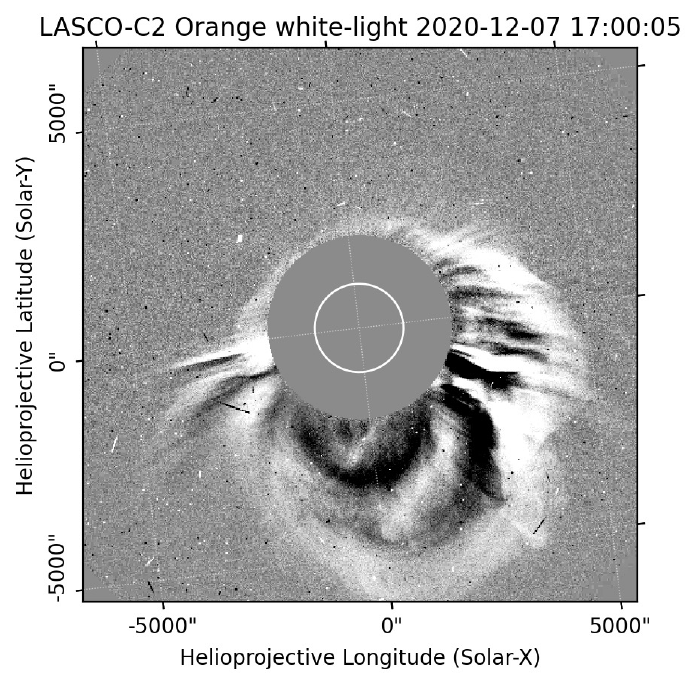} \\
    \includegraphics[width=1.0\linewidth]{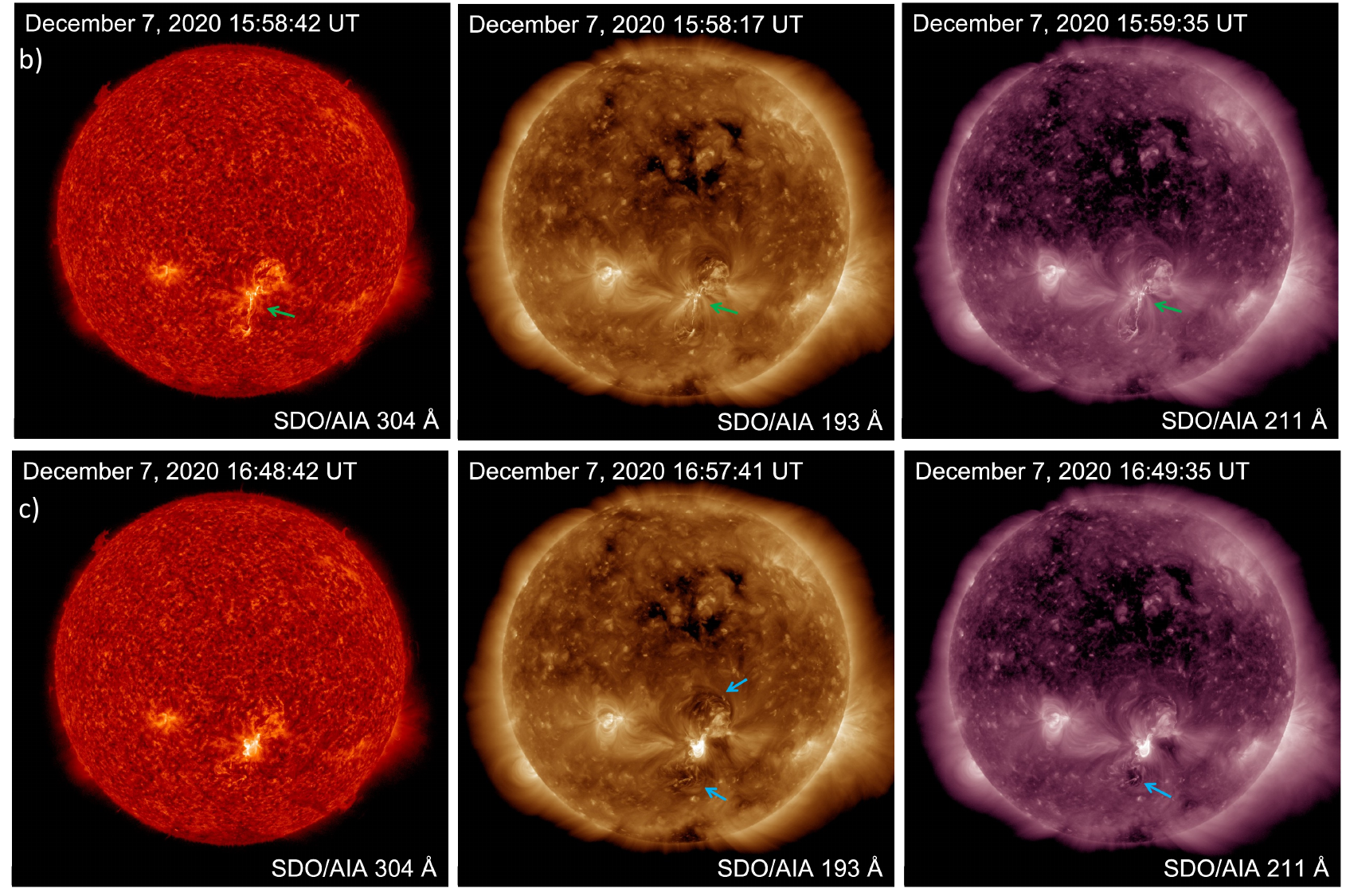}
    \caption {a) Left panel: Profile of the GOES X1.0 flare, showing a typical long duration flare. Middle panel: Magnetic field configuration of the NOAA AR 2790 in HMI magnetogram (adapted from the Solar Monitor). Right panel: SOHO/LASCO~C2 image of CME, obtained with the Pythea tool \citep{Kouloumvakos22}. b) and c) EUV observations by AIA instrument on board the SDO spacecraft at 304, 193, and 211\,\AA\,.~ The images are almost simultaneous and show the development of the erupting filament (marked with green arrows), the flare brightening, and the dimming regions (marked with blue arrows).}
    \label{fig:20201207Combined}
\end{figure*} 

\section{Event study} \label{Events}
Herein, we present the multiwavelength study of two CME-flare events observed from different viewing points. The positions of the spacecraft observing the studied Events are shown for the two events in Figs.~\ref{fig:STEREO_pos}b and c, respectively, as they were obtained from the \href{https://solar-mach.streamlit.app}{Solar-Mach} \citep{Gieseler22} website. The separation angle between STEREO~A and Earth was about 57$^{\circ}$ and 33$^{\circ}$, for Events~1 and 2, respectively.  

\subsection{Event 1}

The first event that we studied was the CME- flare observed on December 7 2020. To our knowledge, no relevant studies of the event have previously been published.  

\subsubsection{CME-flare on December 7 2020} \label{obs1}

At the time of Event~1, only two large numbered ARs were observed on the visible side of the Sun, seen from Earth. The C7.4 GOES flare started at around 15:46~UT, peaked at around 16:32~UT, and lasted for several hours (Fig.~\ref{fig:20201207Combined}a, left panel). The flare, filament, and dimming regions were observed by the SDO/AIA at 304, 193, and 211\,\AA\,, respectively (Fig.~\ref{fig:20201207Combined}b and c). This long-duration flare originated from the NOAA AR 12790 (S23W14), which had at the time of the eruption a very simple, $\alpha$ photospheric magnetic field configuration (Fig.~\ref{fig:20201207Combined}a, middle panel). The second large AR observed at the time of Event~1 was a nearby AR, NOAA AR 12791 (S16E18), which had a $\beta$ configuration of its photospheric magnetic field. The configuration of the global solar magnetic field was rather simple when Event~1 occurred, and it is therefore, together with the source region, mostly the neighbouring NOAA AR 12791 that is influencing the ambient magnetic field and solar plasma through which the studied CME propagates. 

\begin{figure*}[htb]
    \centering
    \includegraphics[width=0.45\linewidth]{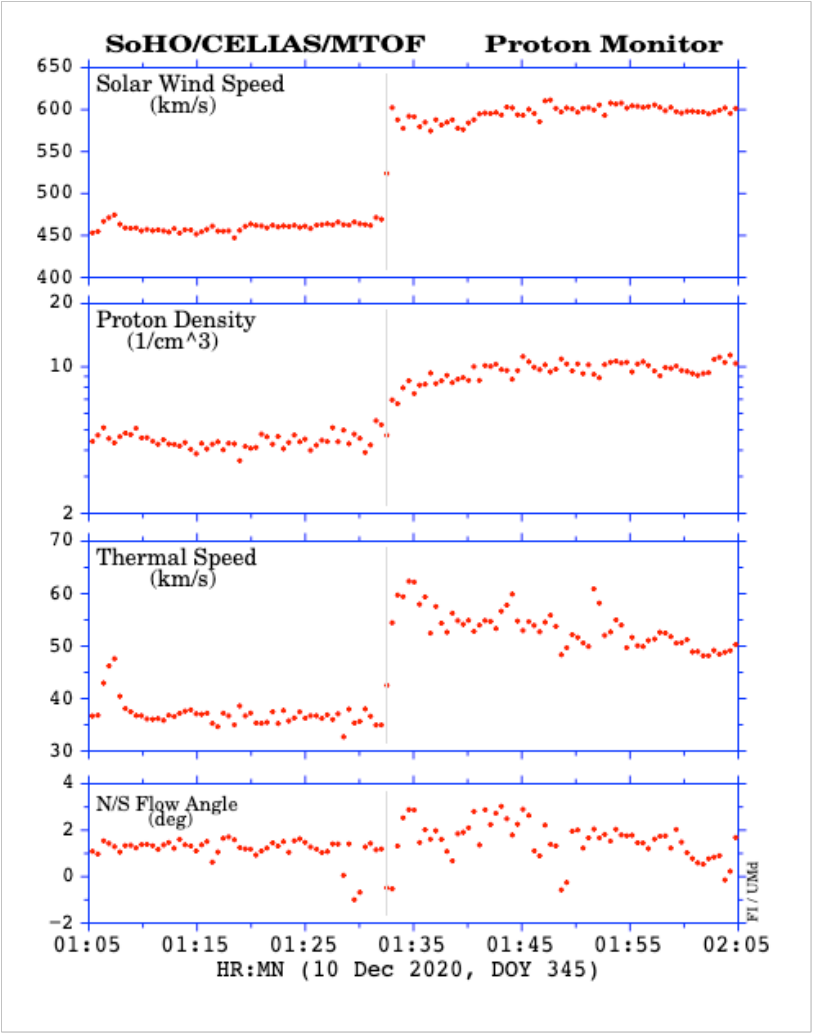}
    \includegraphics[width=0.45\linewidth]{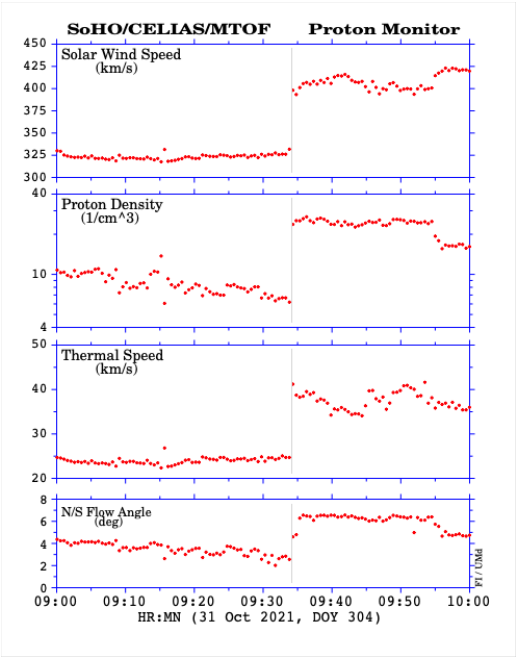} 
    \caption{SOHO/Celias plots that show the FF shocks associated with the two studied CMEs. The left panel shows Event~1 on December 7 2020 and the right one shows the event on October 28 2021. The panels, from top to bottom, show the speed, proton density, thermal speed, and flow angle.}
    \label{fig:Celias}
\end{figure*} 

The studied CME (Fig.~\ref{fig:20201207Combined}c) was first seen in the SOHO/LASCO~C2 field of view at 16:24~UT and in the STEREO~A/COR2 field of view at 15:54~UT. The eruption of the associated filament (marked with green arrows in Fig.~\ref{fig:20201207Combined}b) was observed by the SDO/AIA, as well as the on-disc signatures of the CME; that is, the EIT wave and the coronal dimming (marked with blue arrows in Fig.~\ref{fig:20201207Combined}c). The CME-driven shock wave is clearly visible in the WL coronagraph images by both STEREO~A and SOHO/LASCO (Fig.~\ref{fig:20201207Combined}c). In situ time series at Earth show the arrival of the shock at 01:30~UT on December 10. In Fig.~\ref{fig:Celias}, the shock is indicated by the vertical dotted line, and the left and right panels present Event~1 and Event~2, respectively. 

Although the CME impacted Earth as a flank encounter — only the north CME flank impacted Earth — all four parameters show clear signatures of shock discontinuity, meaning that all four plasma parameters are complying with the shock detection criteria employed in the database of heliospheric shock waves developed by the University of Helsinki \citep{Kilpua15}. Moreover, the shock detection is listed with a 99~\% confidence level in the \href{https://space.umd.edu/pm/figs/figs.html}{SOHO/Celias} in situ shock database. If we employ equations (1) to (4), we obtain for the shock parameters V$_{sh}$=123~km~s$^{-1}$, N$_{sh}$=2.79, T$_{sh}$=2.98, and B$_{sh}$=2.45, and we see that they all comply with the criteria  for a discontinuity to be considered a shock. Moreover, the upstream part of the wave has higher values from the downstream part, for all of the mentioned quantities, and thus the observed shock is identified as a Fast Forward (FF) shock. 

\begin{figure*}[htb]
    \centering
    \includegraphics[width=0.45\linewidth]{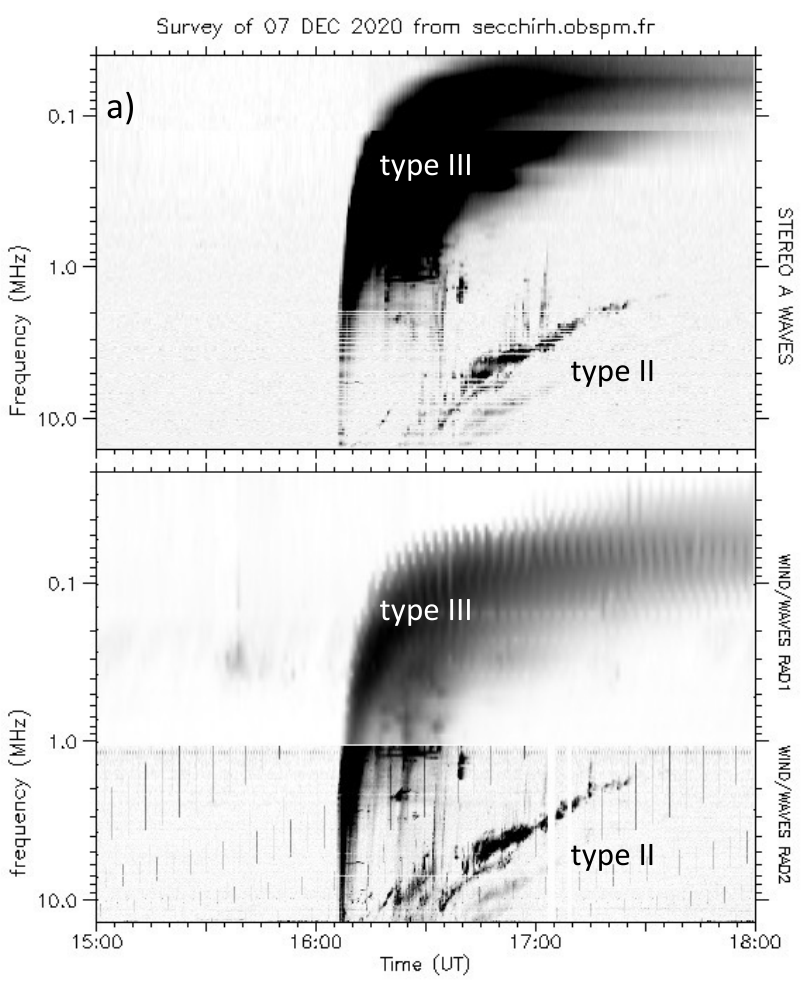}
    \includegraphics[width=0.45\linewidth]{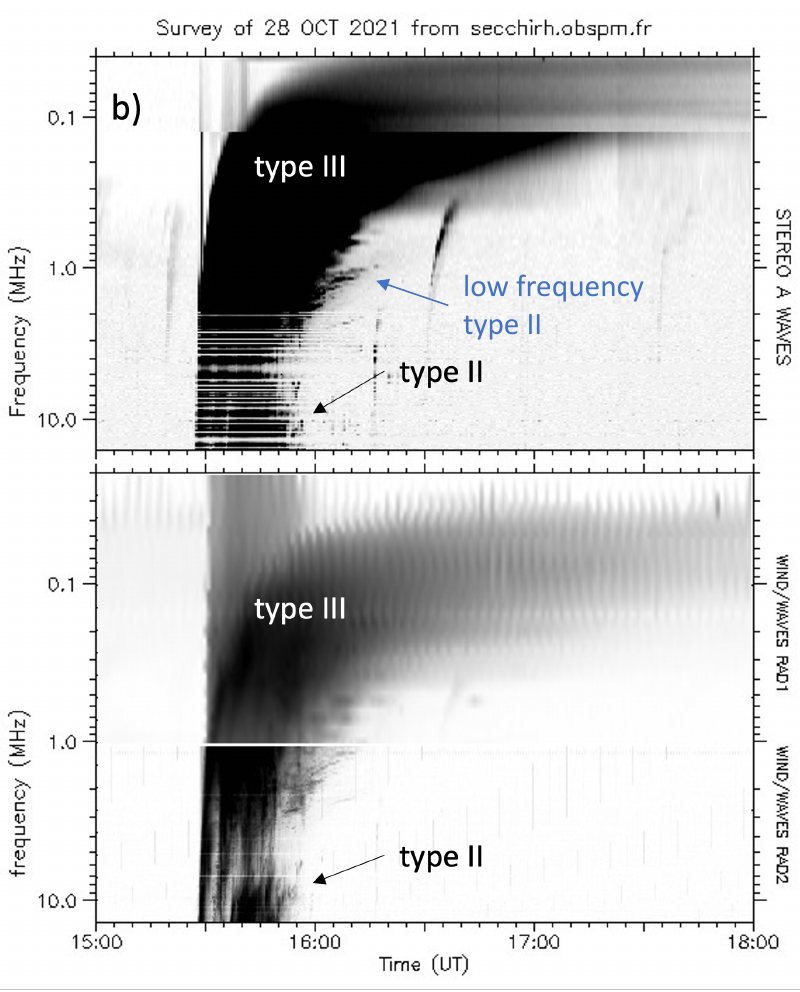} 
    \includegraphics[width=0.45\linewidth]{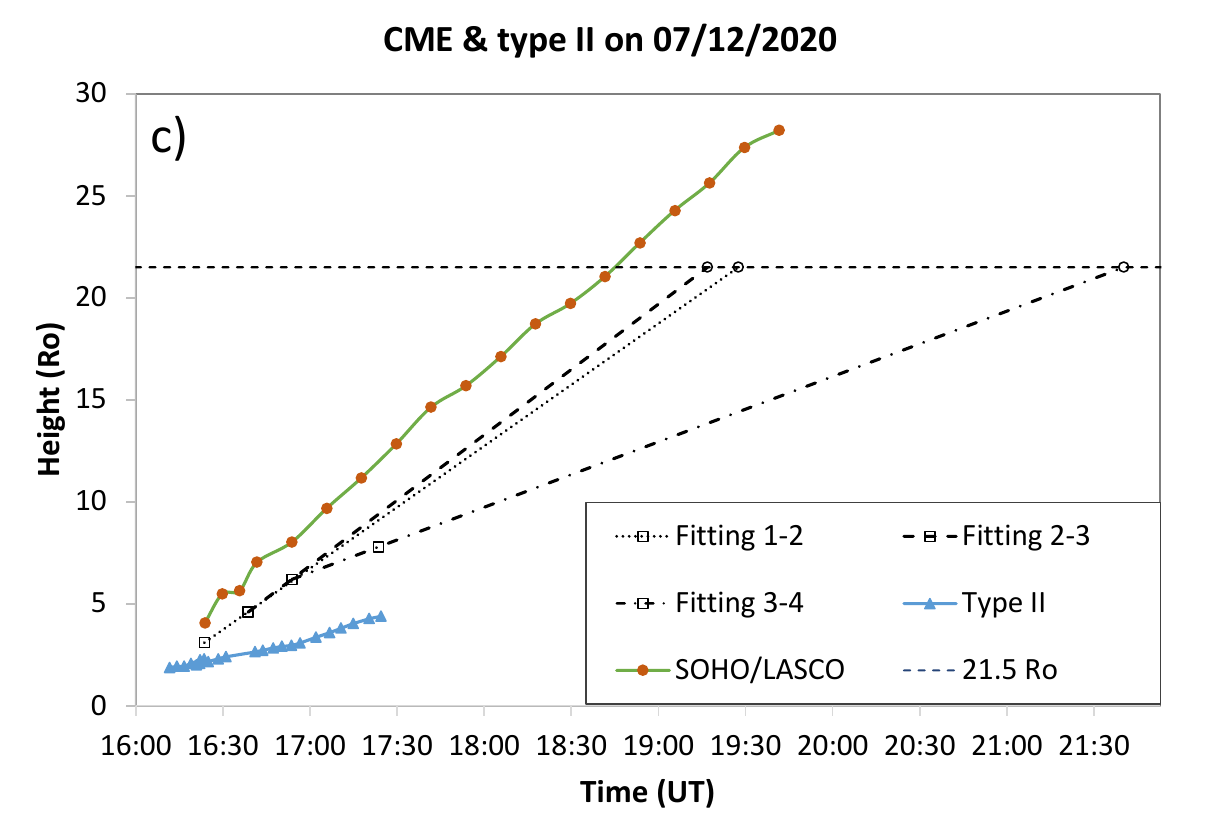}
    \includegraphics[width=0.45\linewidth]{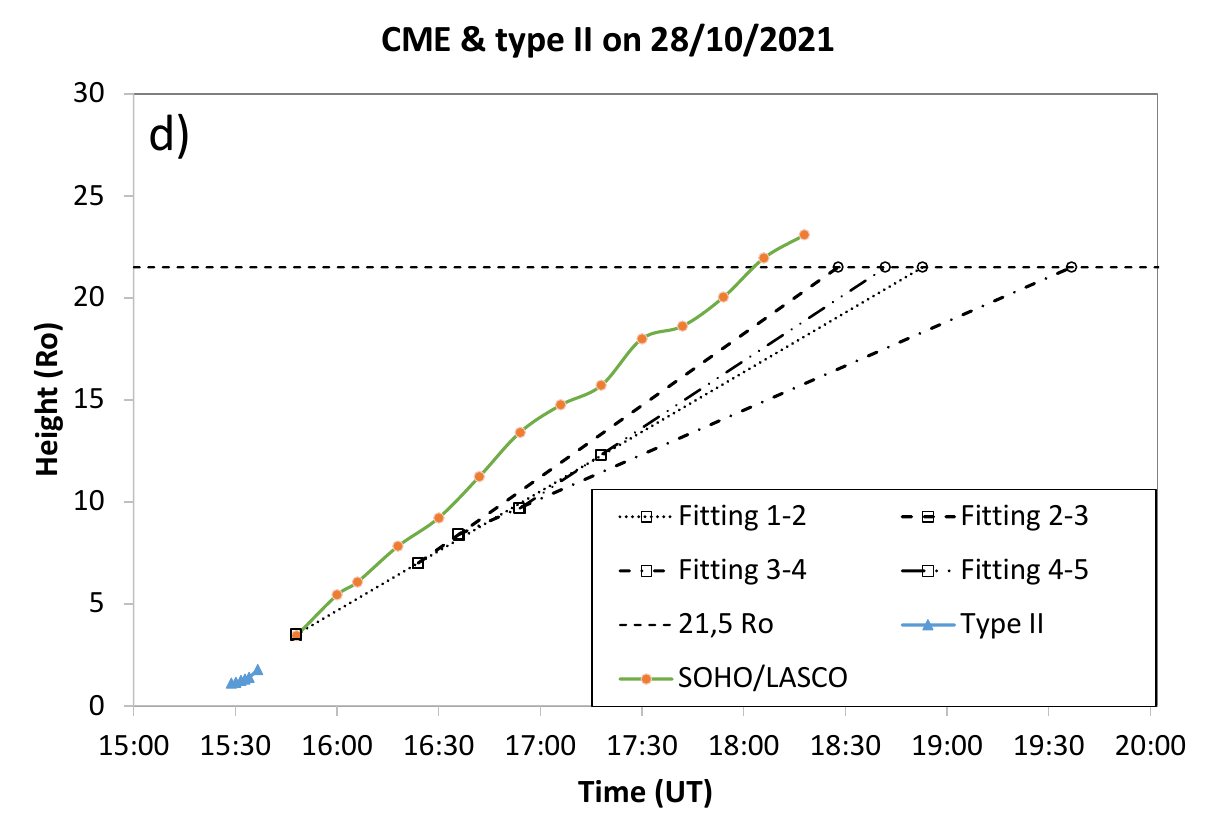}
    \caption{a) and b) Dynamic radio spectra recorded by STEREO A/WAVES (\textit{top}) and WIND/WAVES (\textit{bottom}). In these colour-coded diagrams, dark regions present the enhanced radio emission. a) The radio event associated with the CME-flare observed on December 7 2020 consists of type~III bursts starting mostly at low frequencies and a very prominent type~II burst. b) The radio event on October 28 2021 shows very strong type~III radio emission and a short but intense type~II radio burst that is a continuation of its metric frequency range counterpart. c) and d) The kinematic plots, with info for the speed from the fittings made with GCS, with the black squares, from C2 with the orange points, and from the radio observations with the blue triangles.}
    \label{fig:swaves}
\end{figure*}

The radio observations show interplanetary type~II radio bursts, confirming the existence of the CME-driven shock wave. Dynamic spectra of the radio event, observed by both STEREO/WAVES and WIND/WAVES instruments, are shown in Fig.~\ref{fig:swaves}a, and the analysis of the radio event is presented in Sect.~\ref{radio1}. 

\subsubsection{CME fitting and modelling with EUHFORIA} \label{model1}
\begin{figure*}[htb]
    \centering
    \includegraphics[width=0.31\linewidth]{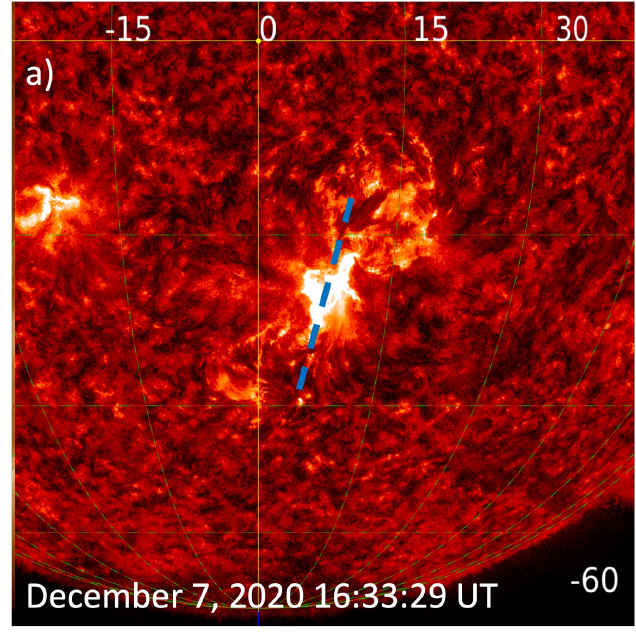}
    \includegraphics[width=0.31\linewidth]{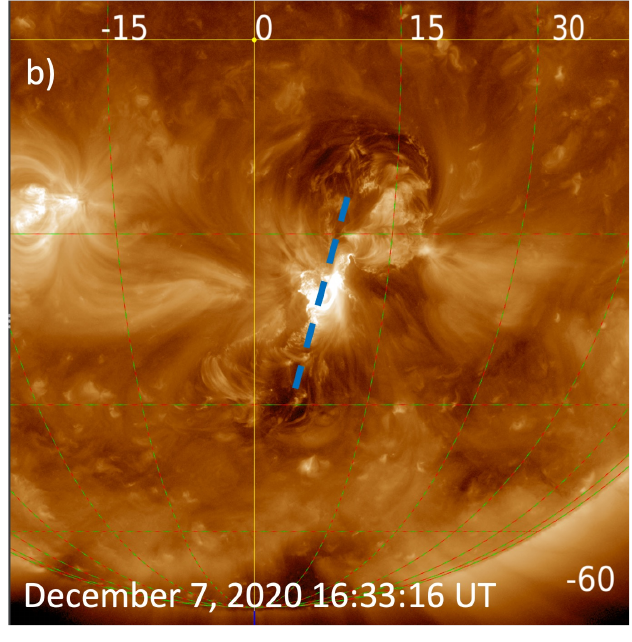}
    \includegraphics[width=0.31\linewidth]{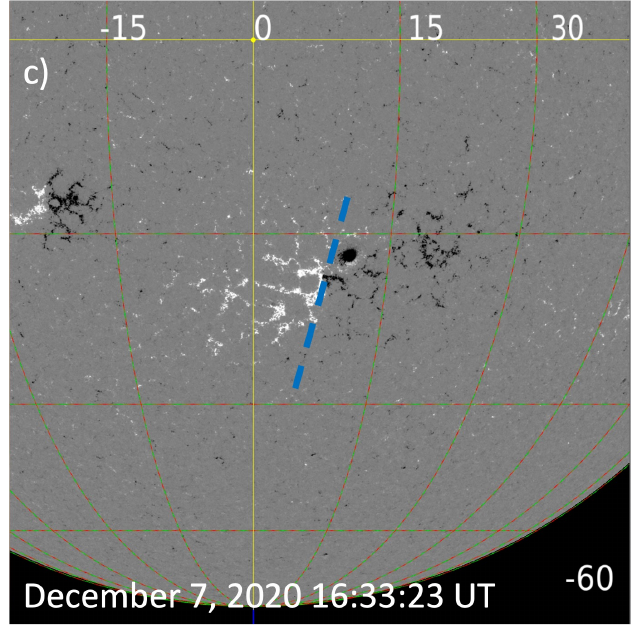}
    \caption{Observations that were used in order to find the orientation of the neutral line of the magnetic loop for Event~1. First, we found it in the magnetogram (c) that is indicated by the dashed blue line, and then we applied it in 304\,\AA\,(a) and 193\,\AA\,(b).}
    \label{fig:20201207_neutralline}
\end{figure*}

The input parameters for the EUHFORIA model were obtained by employing the GCS fitting method (\citealt{Thernisien06}; \citealt{Thernisien09}; \citealt{Thernisien11}). In order to perform the GCS fitting as accurately as possible, we first checked the orientation of the AR neutral line in the SDO/HMI magnetograms (Fig.~\ref{fig:20201207_neutralline}c). We also inspected the orientation of the post-flare loops in the SDO/AIA observations at 304 and 193\,\AA. Figure~\ref{fig:20201207_neutralline} shows the orientation of the main neutral line at 304\,\AA\ at 16:33:29~UT, 193\,\AA\ at 16:33:16~UT, and in the magnetogram at 16:02:38~UT. This analysis allowed us to set up the correct orientation of the legs of the CME grid for the GCS fitting. 

When performing the GCS fittings, we used four pairs of observations from STEREO/COR2 and SOHO/LASCO~C2 and C3, at almost simultaneous times, covering the full radial distance observed by the coronagraphs. The STEREO/COR2 and SOHO/LASCO~C2 pairs of images were named GCS1 to GCS4 (see Table~\ref{table:FittingsDecember}). In such a way we obtained four sets of CME characteristics, which allow us also to inspect the change in the CME propagation direction in the low corona. Table~\ref{table:FittingsDecember} presents the values of the fitting parameters — the latitude, longitude, height, tilt, and half angle — for the different pairs of images. The CME speed was calculated using the height difference and the time difference between two consecutive fittings. The obtained CME speeds were: 840, 1\,160, 1\,240, and 1\,240 km~s$^{-1}$ for GCS1 to GCS4, respectively. These speeds were then used to find the time at which the CME arrives at 0.1~au, and thus the insertion time of the CME in the EUHFORIA model. 

The kinematic of Event~1, using the four different speeds, is presented in Fig.~\ref{fig:swaves}c. The horizontal dashed black curve in both Figs.~\ref{fig:swaves}c and d represents the height of 21.5~R$_\odot$, which is the point at which the CMEs are inserted in EUHFORIA. In these plots, the squares represent the height-time points that we obtained from the GCS fittings. In order to find the time of insertion to EUHFORIA as accurately as possible, we extrapolated each pair of observations up to 21.5~R$_\odot$. The circles on the line of insertion show this time for each pair. Those times are presented in Table~\ref{table:FittingsDecember}. The orange points in Figs.~\ref{fig:swaves}c and d are the height-time measurements of the leading edge of the CME obtained from the \href{https://cdaw.gsfc.nasa.gov/CME_list/index.html}{SOHO/LASCO} CME catalogue. Measurements were obtained from the LASCO~C2 and C3 images on which the 2D projection of the CME was observed, so we named the velocity obtained in such a way a 2D velocity. The blue triangles correspond to the height of the shock wave obtained from the type~II burst and the 1fold Leblanc coronal electron density model (\citealt{Leblanc98}).

To model Event~1, together with the GCS fitting results (Table~\ref{table:FittingsDecember}) we also employed parameters from the \href{https://kauai.ccmc.gsfc.nasa.gov/DONKI/}{DONKI} database. The input parameters from the DONKI database and the result of the GCS4 fitting are shown in Table~\ref{table:Parameters}. The CME parameters from the DONKI database are obtained with two CME analysis tools (CATs): the SWPC\_CAT \citep{Millward13} and the StereoCAT tool, with the first one being the most frequently used. In the catalogue, the CME source location is considered to be the AR associated with the CME, while the location of the CME is the position of the CME as it appears in EUV observations. The CME velocity is derived as a linear fit, to the heights of the height-time points of each fit. 

The reason for selecting two sets of CME parameters from different sources is that the DONKI database is an open access database readily available for the CME modelling, daily updated with new events, and it can be used for the real-time space weather forecasting of the time of the CME's arrival at Earth. In the case of the GCS technique, though, we obtained the CME parameters by applying the fitting procedure ourselves, being as accurate as possible and using different types of observations that might not be real-time available. Employing these two sets of  parameters allows us also to compare the accuracy of the almost real-time modelling and the post-event analyses.

Before performing the CME modelling, we first optimised the solar wind modelling (Sect.~\ref{methods}). For Event~1, we used three GONG synoptic magnetogram maps on December 7 2020 at different times (04:00, 10:00, and 13:00~UT) in order to find the one that would give the modelling results in best agreement with the observed solar wind in situ data. The one that provided the best result in the solar wind modelling with EUHFORIA was the magnetogram at 10:00~UT. This magnetogram was used to model the background solar wind in which we inserted the CME.  

\begin{figure*}[htb]
    \centering
    \includegraphics[width=1.0\linewidth]{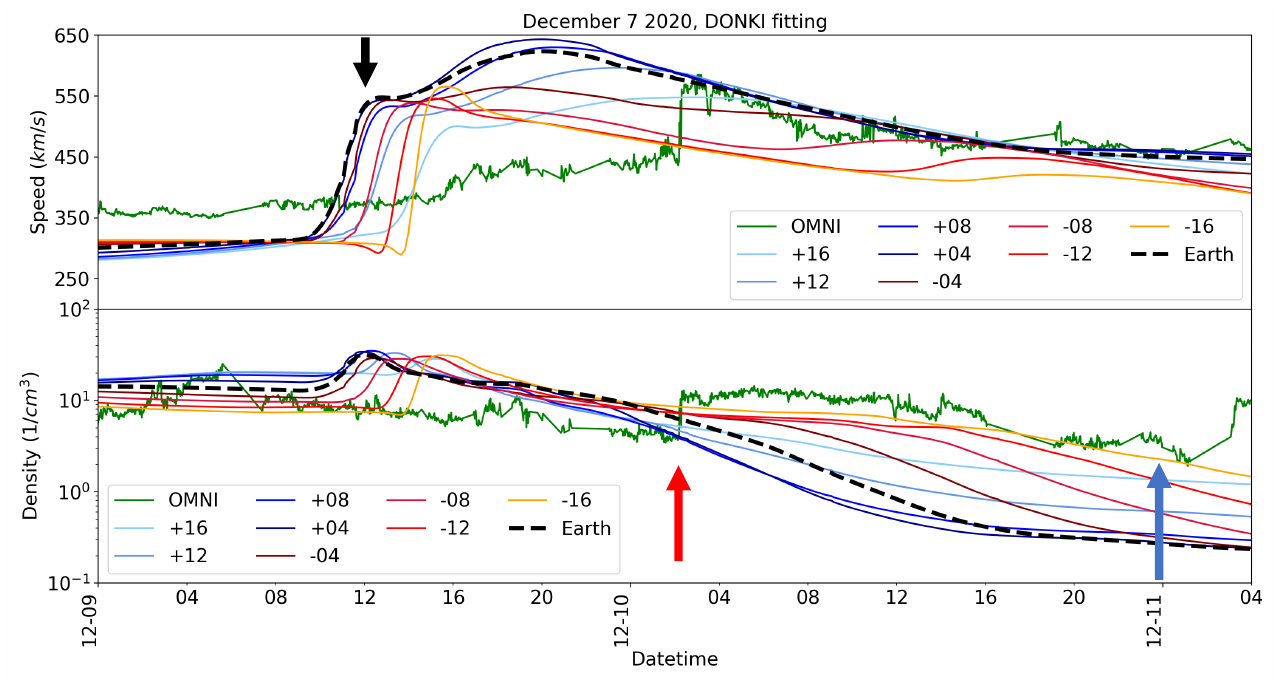}
    \caption{Comparison of the in situ observations at Earth and the modelled velocity (top panel) and density time series (bottom panel) obtained in EUHFORIA simulations for Event~1. The in situ data are presented as the green line and the modelled time series have different shades according to the position at which they were estimated (blue shades are at the positions above Earth and red shades are time series bellow Earth, see also Fig.~\ref{fig:STEREO_pos}a). The modelling results were obtained using the CME characteristics from the DONKI database. The red arrow indicates the observed arrival time at Earth. The black arrow indicates the modelled arrival time of the disturbance to Earth (dashed black curve) at zero degree latitude and longitude. The blue arrow indicates the estimated arrival time from the measurements of the radio observations. The modelled arrival time at Earth is significantly earlier than the observed one with $\Delta$t~$\approx$~-14h (estimated as a difference between the in situ shock arrival and the middle shock amplitude).}
    \label{fig:December_DONKI}
\end{figure*}

After optimising the background solar wind modelling with EUHFORIA, we modelled the CME using first the input sets of the CME characteristics obtained from the DONKI database. The in situ time series of the solar wind plasma parameters at Earth, together with the results of the CME modelling with the DONKI parameters, are shown in Fig.~\ref{fig:December_DONKI}. The solar wind speed and density are presented in the top and bottom panels, respectively. The green curve shows the dynamics of the solar wind in situ time series obtained from \href{https://omniweb.gsfc.nasa.gov}{OMNI} database. The dashed black curve presents the time series of modelling results at the position of Earth. The curves in shades of blue and red correspond to the modelling results at the positions of virtual spacecraft located above and below Earth, respectively (Fig.~\ref{fig:STEREO_pos}a). The observed arrival of the CME at Earth is indicated by the red arrow and the modelled arrival time is marked by the black arrow. 

\begin{figure*}[htb]
    \centering
    \includegraphics[width=1.0\linewidth]{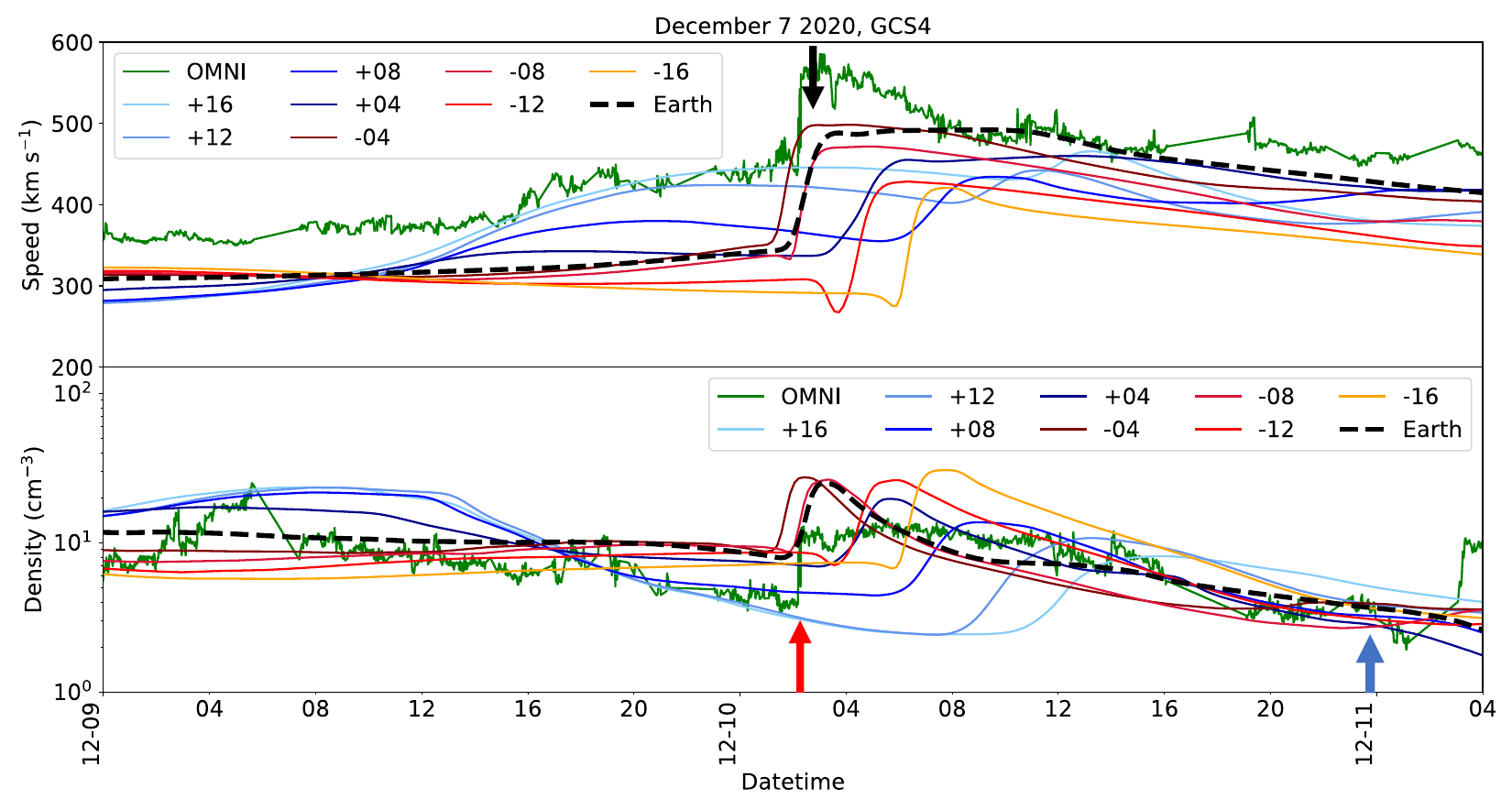}
    \caption{Comparison of the in situ observations at Earth and the modelled velocity (top panel) and density time series (bottom panel) obtained in EUHFORIA simulations. The in situ data are presented as the green line and the modelled time series have different shades according to the position at which they were estimated (blue shades are at the positions above Earth and red shades are time series below Earth). The modelling results were derived using the CME characteristics obtained from the best GCS fit, e.g. GCS4, at which the CME was the furthest away from the solar surface. The red arrow indicates the observed arrival time at Earth. The black arrow indicates the modelled arrival time of the shock wave to Earth (dashed black curve) at zero degree latitude and longitude. The blue arrow indicates the estimated arrival time from the measurements of the radio observations. The modelled time of the shock wave's arrival at Earth is almost at the same time as the observed one, $\Delta$t~$\approx$~+20min. The observed $\Delta$t is different for different latitudes ranging from 10--14~h.}
    \label{fig:GCS-fit4}
\end{figure*}

We note that a significant difference between the observed and the modelled shock arrival of about $\Delta$t~$\approx$~-15~h was obtained (the negative sign indicates an earlier arrival). The existence of the double peak of the velocity profile and two-times-larger modelled shock wave amplitude was induced by the high-speed stream and the particularity of the DONKI parameters (see also discussion about Fig. \ref{fig:GCS-fit4}). Modelling results indicate that the CME parameters obtained from the DONKI database do not provide very accurate results for the studied CME. It is clear that at least the CME velocity obtained from the DONKI database and input at 0.1 au in EUHFORIA was higher than the real CME velocity, but also other input parameters were possibly different than those the CME most likely has. The observed $\Delta$t is also different for different latitudes above and below Earth (blue and red shades, respectively) with the range of inaccuracy being about 10~--~15h. The arrival times estimated at the time series of virtual spacecraft at latitudes strongly above and below Earth (+16$^{\circ}$ and -16$^{\circ}$, respectively), which are more CME flank impacts, are more accurate; that is, closer to the observed time. The blue curve at +16$^{\circ}$ shows a significantly less steep time profile than the other blue and red curves, indicating that although the studied CME is Earth-directed its main propagation direction is oriented slightly southward from the Sun-Earth line. Also, it is important to mention that the dips that appear before the shock, in some of the modelled curves, are an artefact of the code, with no physical meaning. Same dips appear also in Figs.~\ref{fig:GCS-fit4}, \ref{fig:Donki_Orig}, and \ref{fig:GCS-44}. 

In the next step of the study, we performed the GCS fitting of the CME. For the 3D fitting process, we used the four pairs of coronagraph observations, GCS1 to GCS4. The parameters presented in Table~\ref{table:FittingsDecember}, except from the height and the tilt, were used as input for the EUHFORIA runs with the cone model. In the fitting, the values of the longitude and the latitude represent the location of the CME nose. The fact that these parameters change between the fittings means that the CME experiences deflections and rotations as it propagates away from the Sun. Those changes in the CME direction also affect the estimation of the CME speed. In order to demonstrate the orientation of the flux rope in the GCS fittings, we provide an example of them in Fig.~\ref{fig:Example_fit_Event1} in the appendix. 

In order to understand how much variation different CME parameters experienced, we also calculated the average and the standard deviation values for the latitude, the longitude, the half angle, and the tilt (last two columns of Table~\ref{table:FittingsDecember}). The standard deviation, which shows the spreads from the mean value, is smallest for the longitude, but not negligible for the other parameters, being about 10~\% for latitude and half angle and 6~\% for the tilt. This is an indication that the CME is indeed rotating (change in the tilt) and changing its propagation path, in the early stages of its evolution. The change in both the longitude and the latitude indicates that the CME experiences deflections that shift its direction of propagation, which is in this event towards the north. The change in the half angle indicates that the CME also expands during its propagation. Although these changes are not too large, they are influencing the propagation direction of the CME anyway. The change in the CME direction was also observed at lower heights and in the EUV data, but as this was more difficult to quantify and as it does not directly influence the estimation of the CME parameters using the GCS fitting (starting at larger heights) we do not discuss it here. 
The CME speed obtained from the GCS fitting was estimated by using the difference in height between the two fitted CME positions, and the time difference between those two points. For the first velocity estimation (GCS1) we also used the 2D information from the SDO/AIA and STEREO/COR1 observations.

\begin{table*}
    \caption{Parameters from the GCS fittings for the CME simulations for Event~1.} 
    \centering
    \begin{tabularx}{0.95\linewidth}{@{\extracolsep{\fill}}|l|r|r|r|r|r|r|} 
    \hline
    \multicolumn{7}{|c|}{December 7, 2020} \\
    \hline
    Parameters & GCS1 & GCS2 & GCS3 & GCS4 & Mean & Stdev \\ 
    \hline\hline
    Time in LASCO (UT) & 16:24:08 & 16:36:07 & 17:00:07 & 17:30:07 & - & - \\
    \hline 
    Time in STA (UT)& 16:24:00 & 16:39:00 & 16:54:00 & 17:24:00 & - & - \\
    \hline 
    Time at 0.1\,AU (UT) & 20:37:55 & 19:28:00 & 19:17:00 & 19:32:00 & - & - \\ 
    \hline
    Latitude ($^{\circ}$)   & -29.4 & -29.4 & -24.4 & -24.6 & 26.95 & 2.83 \\ 
    \hline
    Longitude ($^{\circ}$)  & 18.3 & 17.7 & 17.7 & 18.3 & 18.00 & 0.35 \\ 
    \hline
    Height (R$_{\odot}$)    & 3.1 & 4.1 & 6.2 & 7.8 & - & - \\
    \hline 
    Half Angle ($^{\circ}$) & 24.1 & 24.1 & 24.8 & 29.4 & 25.60 & 2.55 \\
    \hline 
    Tilt                    & 57.3 & 57.0 & 56.5 & 50.0 & 55.20 & 3.48 \\
    \hline
    Speed (km~s$^{-1}$)            & 840 & 1160 & 1240 & 1240 & - & - \\ 
    \hline 
    Day at 1 AU & December 10 & December 10 & December 10 & December 10 & - & - \\ 
    \hline 
    Time at 1 AU (UT) & 22:13:30 & 13:33:40 & 06:33:40 & 02:23:40 & - & - \\ 
    \hline
    \end{tabularx} 
    \label{table:FittingsDecember}
\end{table*}

Figure~\ref{fig:GCS-fit4} presents the results of the EUHFORIA simulations employing the last set of the GCS parameters (GCS4), when the CME was at a height of about 7.8~R$_{\odot}$. The panels are in the same order and the curves and the arrows represent the same quantities as in Fig.~\ref{fig:December_DONKI}. The GCS4 provided the best agreement between the modelled and observed times of the shock wave arrival at 1~au, with the difference being only about $\Delta$t~$\approx$~+10 minutes (Fig.~\ref{fig:GCS-fit4}). We believe that this accurate modelling result arises from the point that at a height of 7.8~R$_{\odot}$ the modelled CME has experienced the majority of the deflections, rotation, and the change in the propagation direction, and it propagates away from the Sun mostly radially.  

In the case of the run with the DONKI parameters, Fig.~\ref{fig:December_DONKI}, we see that, apart from the fact that the modelled arrival time of the CME is much earlier than the observed one, the virtual spacecraft at the position of Earth is the first to record it, along with the one that was located 4$^{\circ}$ above it. From that, we can assume that the CME modelling results with the DONKI parameters indicate a direct impact of the CME at Earth. 

The curves in shades of blue and red (Fig.~\ref{fig:GCS-fit4}) correspond to the modelling results at the positions of virtual spacecraft located above and below Earth, respectively (Fig.~\ref{fig:STEREO_pos}a). The time series modelled by EUHFORIA at latitudes below and above Earth show at first hand somewhat chaotic behaviour. The time series at virtual spacecraft located above Earth (blue shades) reflects the interaction of the CME with the preceding fast solar wind originating from the negative polarity coronal hole (Fig.~\ref{fig:20201207Combined}b and c, right panels). That fast solar wind first impacted the region above Earth. All of the above-Earth time series show an increase in the solar wind velocity starting from about 10:00~UT on December 9 2020, which is prior to the CME's arrival at Earth. The solar wind increase starts first in the time series mapped at the highest latitudes (+16$^{\circ}$) and this is followed by the lower latitudes' time series. The solar wind density decreases as the solar wind velocity increases, which is typical for the solar wind originating from coronal holes. The solar wind time series at Earth does not show such a behaviour; neither does the time series below Earth (red shades). 

The behaviour of the time series below and above Earth (Fig.~\ref{fig:GCS-fit4}) is not the same as the modelling results when the DONKI parameters were considered (Fig.~\ref{fig:December_DONKI}). This difference appears because the CME modelled with the DONKI parameters arrives significantly earlier (15 hours difference with GCS4), and does not interact with the fast solar wind in the same way as the CME modelled with GCS4 input parameters. The impact of the CME with DONKI parameters was more of a direct hit than that of the one with GCS4 parameters, which shows the first arrival of the shock at a latitude of -4$^{\circ}$, indicating that the CME was propagating somewhat southward from the Sun-Earth line. We note that the difference between the CME latitude source region is also -4$^{\circ}$. This result can also be expected because the CME half-width with DONKI parameters is 11.6 degrees larger than our best fit, and longitude and latitude are closer to the center of the solar disc, ensuring a more direct CME impact (for a comparison of the difference between the CME input values from the DONKI webpage and our best modelled fit, see Table \ref{table:Parameters}). 

\begin{figure*}[htb]
    \centering
    \includegraphics[width=1.0\linewidth]{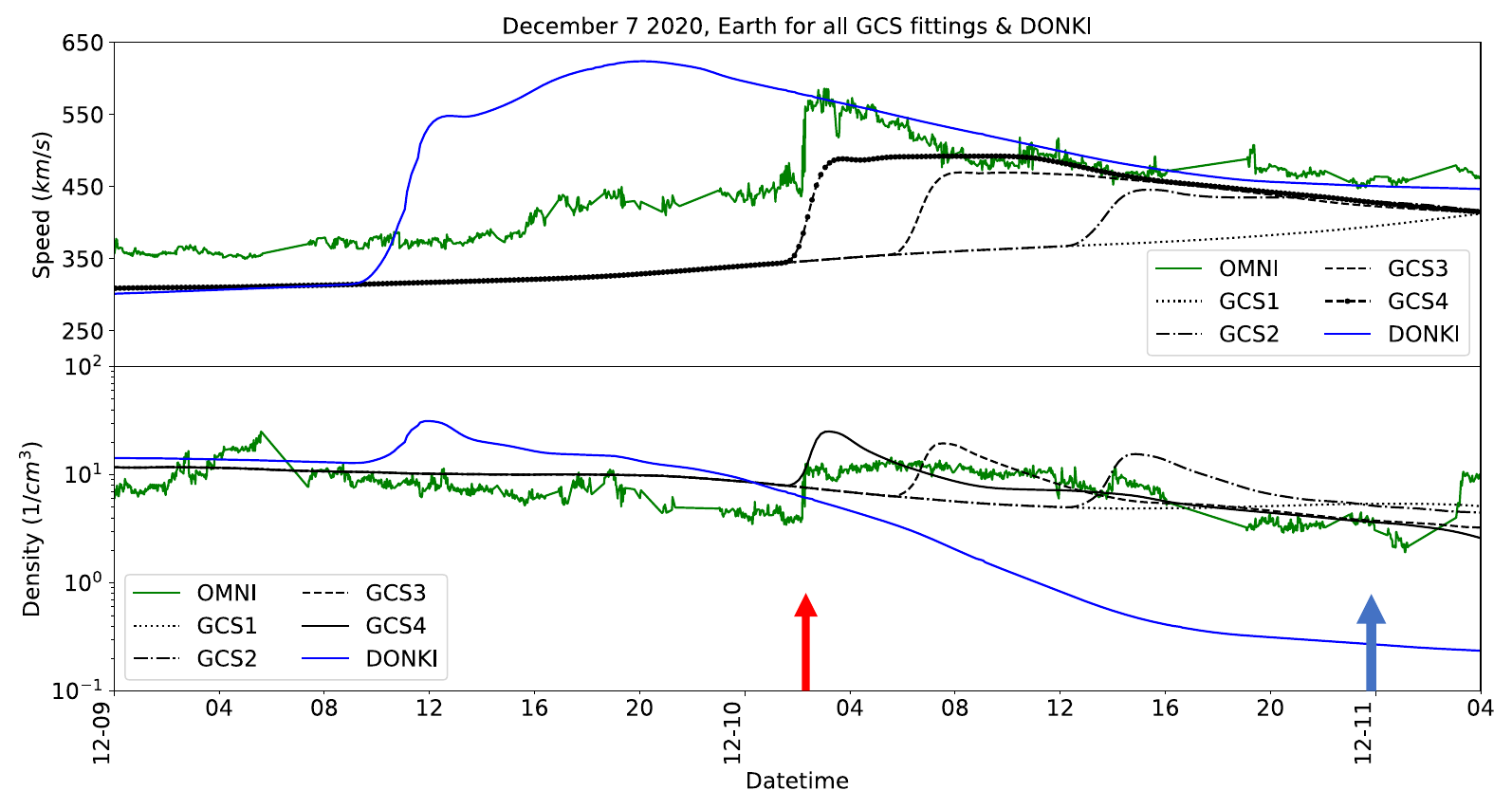}
    \caption{Comparison of the in situ observations at Earth and the modelled velocity (top panel) and density time series (bottom panel) obtained in EUHFORIA simulations, for the cases of Earth, at zero degrees longitude and latitude, in the case of Event~1. The in situ data are presented as the green line and the modelled time series have different line styles according to the fitting that they present. We also include with the blue curve the results for Earth, for the case of the DONKI parameters. The modelling results were obtained using the CME characteristics obtained from the GCS fittings (Sect.~\ref{model1}). The red arrow indicates the observed arrival time of the disturbance at Earth. The blue arrow indicates the estimated arrival time from the measurements of the radio observations. The majority of the modelled Earth arrival times are late in comparison with observations, but they gradually shift towards the observed time as we go from GCS1 up to GCS4 and reconstruct the CME at larger heights. The observed $\Delta$t is different for different latitudes, ranging from 20~mins to 20~h.}
    \label{fig:December_earth_all}
\end{figure*} 
 
The comparison of the modelled Earth arrival time when employing GCS1, GCS2, GCS3, and GCS4 (Table~\ref{table:FittingsDecember}) shows that the modelled arrival time of the disturbance constantly improves with the CME input parameters obtained from the GCS fitting at subsequently increasing heights (Fig.~\ref{fig:December_earth_all}). The fitting at the greatest heights provides the modelling results closest to the observed one. It is important to note that not only the arrival time is improved but also the magnitude and the shape; the steepness of the shock is constantly increasing from GCS1 towards GCS4. This is because the differences in the CME input parameters and the CME propagation direction will also determine which part of the CME — that is, the flanks or the leading edge — will impact Earth and also be detected in the in situ observations at Earth. A recent study by \cite{Palmerio24} shows that CME signatures in the in situ observations, by even very closely positioned spacecraft, can show very different characteristics. This behaviour is possibly also reflected in our modelling results. 

A number of studies (e.g. \citealt{Gui11}; \citealt{Rollett14}) have shown that the CMEs in general experience deflections at low heights and therefore do not propagate fully radially outwards up to 0.1~au. The CME deflections can be due to the way the CME erupts or due to the proximity of other coronal structures, such as streamers or coronal holes, which can then interact with the CME. Such interactions can affect the propagation direction of the CME in its early stages of evolution, and thus also its arrival time at Earth. On the other hand, if we use observations of the CME at larger heights so that the derived values for the CME's characteristics are after the CME has undergone the deflections and major changes in the propagation direction, we can expect more accurate modelling results. 

\begin{figure*}[htb]
    \centering
    \includegraphics[width=0.32\linewidth]{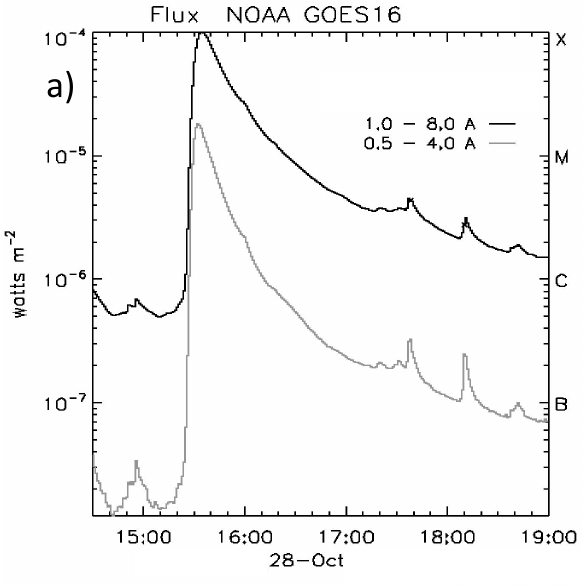}
    \includegraphics[width=0.32\linewidth]{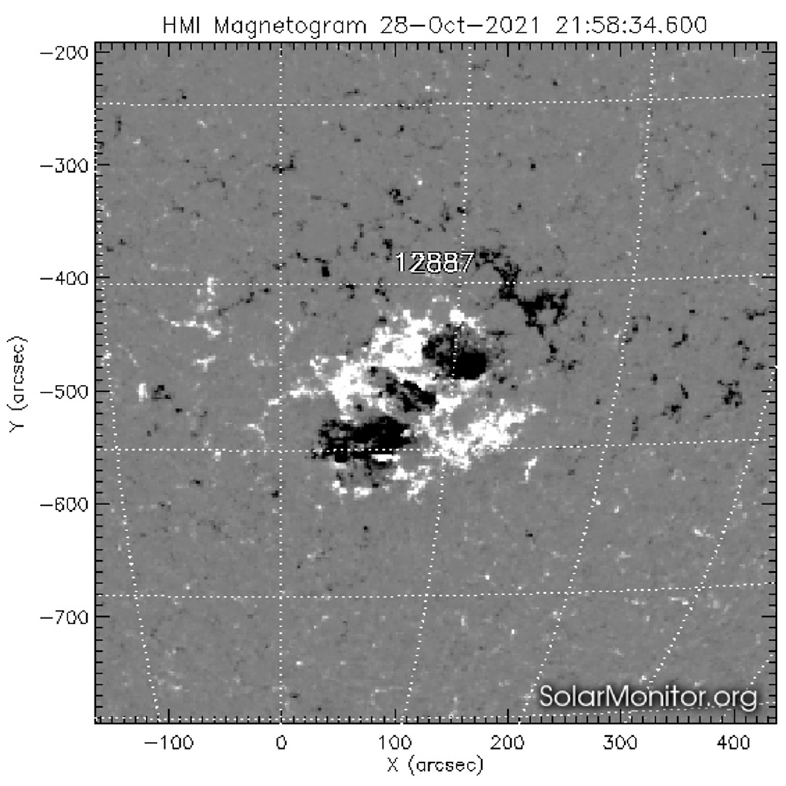}
    \includegraphics[width=0.33\linewidth]{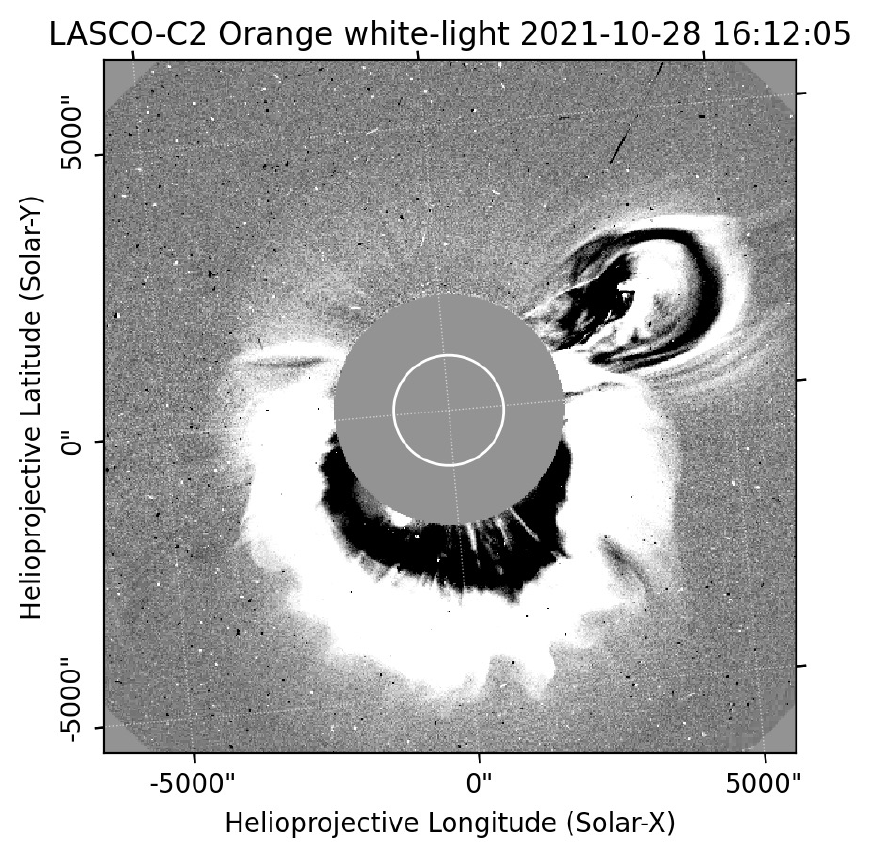} \\
    \includegraphics[width=0.95\linewidth]{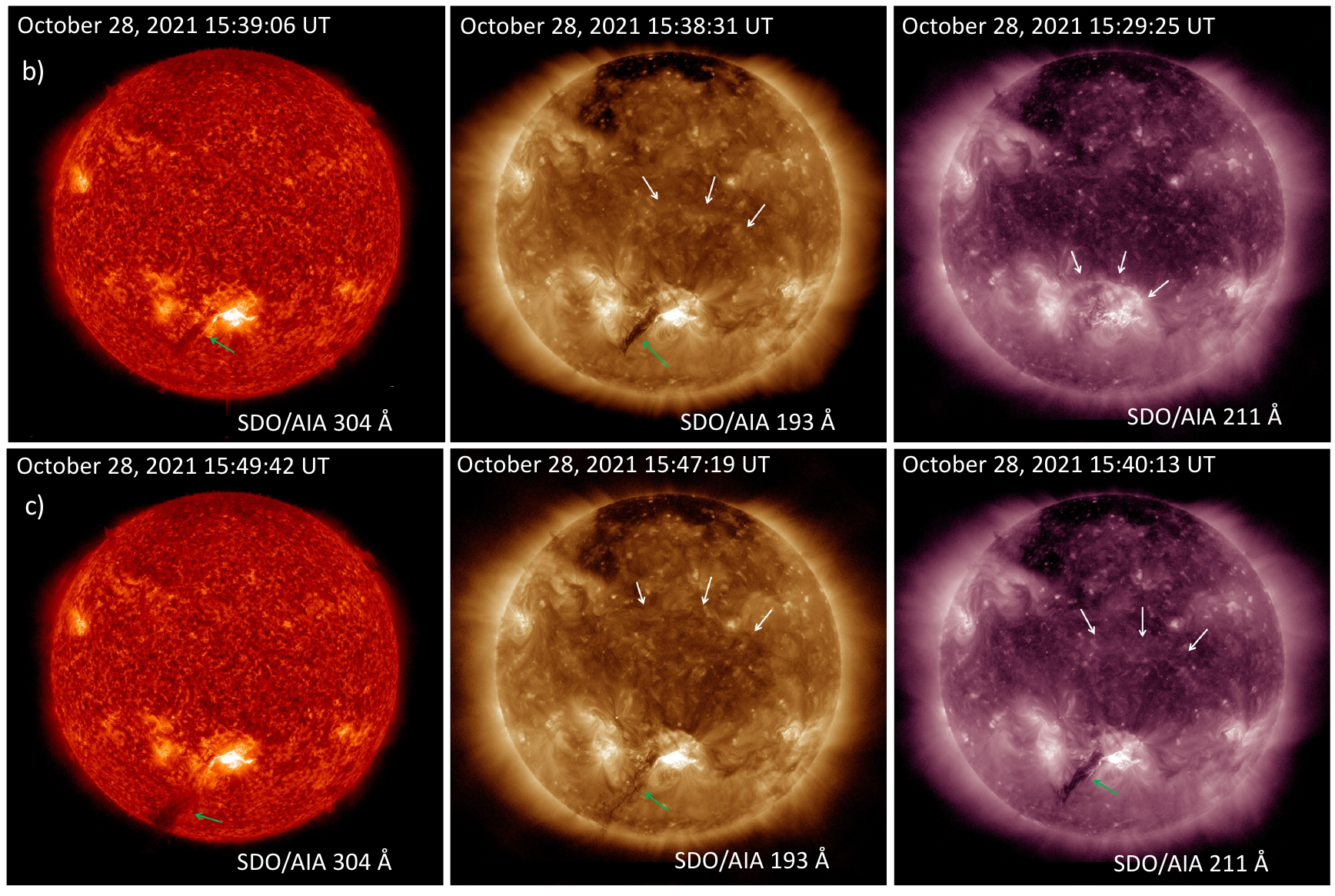}
    \caption {a) Left panel: Profile of the GOES X1.0 flare showing a typical long-duration flare. Middle panel: Magnetic field configuration of the NOAA AR 2887 observed in the HMI magnetogram. Right panel: SOHO/LASCO~C2 image of CME (obtained with the PyThea tool). b) and c) EUV observations from the AIA instrument on board the SDO spacecraft. The first column shows the observations at 304~\AA; the second and third columns show observations at 193~\AA\, and the 211~\AA\,, respectively. The times of the considered images are different in order to better capture the development of the eruption and the associated wave.}
    \label{fig:GOESflare}
\end{figure*} 

\subsubsection{RADIO observations} \label{radio1}
The radio event associated with the CME on December 7 2020 consisted of type~II and type~III radio bursts (Fig.~\ref{fig:swaves}a). The small group of type~III radio bursts was observed starting at about 16:05~UT and was followed by type~II radio burst shock wave signatures. The type~II radio burst started at around 16:35~UT and ended at around 17:30~UT, as is seen from the STEREO~A/WAVES observations, and was observed in WIND/WAVES data during the time interval between 16:20~and~17:30~UT. The ending frequency of type~II was approximately 2~MHz in both STEREO and WAVES observations. Both instruments also show clear fundamental and harmonic bands of type~II being split into two quasi-parallel emission lanes, the so-called band split lanes that are often considered to be emissions from the upstream and downstream shock region (\citealt{Smerd74}; \citealt{Smerd75}; \citealt{Vrsnak01}; \citealt{Magdalenic02}). The type~II burst is well observed in both STEREO and WAVES dynamic spectra, which indicates that its source is located in 3D space somewhere in between the two spacecraft. Since the radio emission is more intense and lasts longer in the WIND/WAVES observations, we can deduce that its direction of propagation was somewhat closer to Earth (\citealt{Magdalenic14}, \citealt{Jebaraj20}) than to STEREO~A. 

\begin{figure*}[htb]
    \centering
    \includegraphics[width=0.31\linewidth]{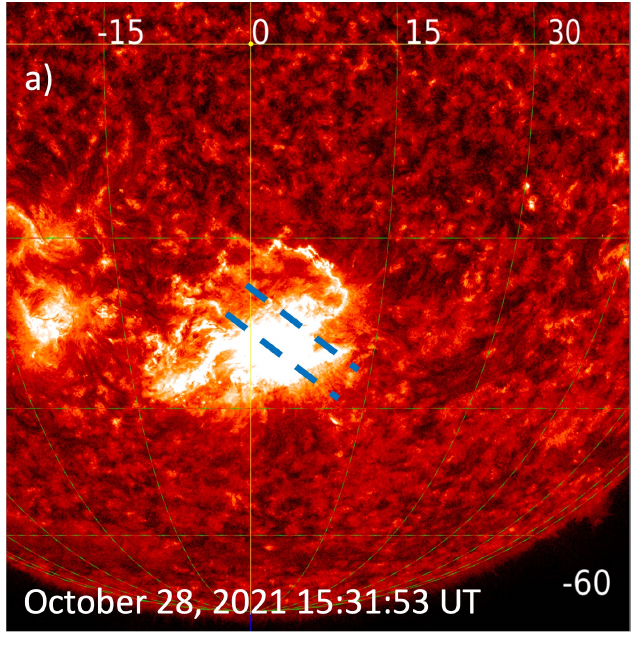}
    \includegraphics[width=0.31\linewidth]{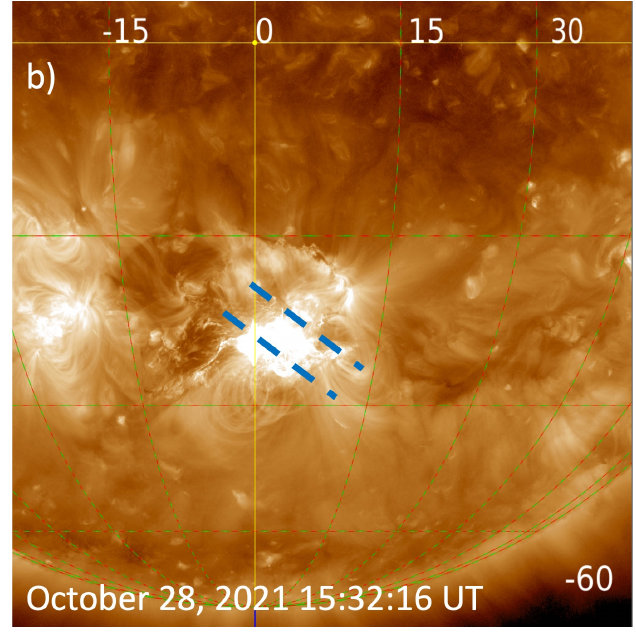}
    \includegraphics[width=0.31\linewidth]{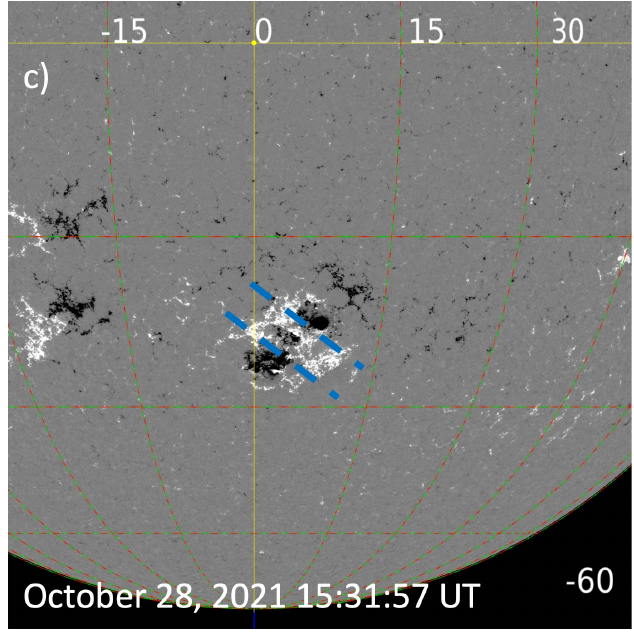}
    \caption{Observations used to find the orientation of the AR neutral line, and therefore also the flux rope orientation for Event~2. First, we identified the neutral line, indicated by the two dashed blue lines, in the magnetogram (panel c), and then we compared it with the post-eruption arcade in 193~\AA\ (panel b), and positioned it in the 304~\AA\ image for comparison (panel a).}
    \label{fig:20211028_neutralline}
\end{figure*} 

Employing the frequency drift of the type~II burst and the 1fold Leblanc electron density model (\citealt{Leblanc98}), we estimated the shock wave heights (Fig.~\ref{fig:Celias}), and the velocity of the CME-driven shock. The velocity estimated from the radio observations taken by both WIND and STEREO~A was found to be about 520~km~s$^{-1}$, which gives the shock arrival time to Earth as around 23:50~UT on December 10 (marked with the blue arrow in Figs.~\ref{fig:December_DONKI}, \ref{fig:GCS-fit4} and \ref{fig:December_earth_all}). If we now compare the arrival time of the shock wave estimated from the radio observations, modelling results with EUHFORIA, and in situ observations (Fig.~\ref{fig:GCS-fit4}) we see that the radio observations give $\Delta$t~$\approx$~+21~h. Such a large difference between the estimated shock arrival time and the observed shock arrival time at 1~au indicates that the type~II emission was possibly generated at the CME flank closer to the southern polar coronal hole, and not close to the CME's leading edge. In order to verify this conclusion, the positioning of the type~II radio sources in 3D space would be necessary (see e.g. \citealt{Magdalenic14}, \citealt{Jebaraj20}). 

Another way to estimate the arrival of the shock at Earth is by using the speed that is given in the \href{https://cdaw.gsfc.nasa.gov/CME_list/index.html}{SOHO/LASCO} catalogue. This speed is estimated by employing the CME height-time measurements performed using the SOHO/LASCO images (2D information) and this is why we named it the 2D speed. 

\begin{table*}
    \caption{Parameters used for the CME simulations for Event~1 and Event~2 obtained from the DONKI database and from the last GCS fitting in the SOHO/LASCO~C2 field of view. For the density we assumed the value of $\rho_{CME}$~=~10$^{-18}$~K and for the temperature the value of T$_{CME}$~=~0.8~MK.}
    \centering
    \begin{tabularx}{0.95\linewidth}{@{\extracolsep{\fill}}|l|r|r|r|r|r|} 
    \hline
    - & \multicolumn{2}{c|}{December 7 2020} & \multicolumn{3}{c|}{October 28 2021} \\
    \hline
    Method & DONKI & GCS4 & DONKI & GCS4 & GCS5 \\ 
    \hline\hline
    Time in LASCO (UT)& - & 17:24:00 & - & 16:54:07 & 17:18:07 \\
    \hline 
    Time in STA (UT)& - & 17:24:00 & - & 16:53:30 & 17:23:30 \\
    \hline 
    Time at 0.1~au (UT) & 18:49:00 & 19:32:00 & 18:52:00 & 18:58:00 & 18:45:00 \\ 
    \hline
    Latitude ($^{\circ}$) & -21 & -24.6 & -17 & -42 & -44 \\ 
    \hline
    Longitude ($^{\circ}$) & 12 & 18.3 & 0 & 0 & 0 \\ 
    \hline
    Height (R$_{\odot}$) & - & 7.8 & - & 9.7 & 12.3 \\
    \hline 
    Half Angle ($^{\circ}$) & 41 & 29.4 & 49 & 49 & 49 \\
    \hline 
    Speed (km~s$^{-1}$) & 1383 & 1236 & 1270 & 1110 & 1110 \\ 
    \hline 
    Day at 1~au & December 9 & December 10 & October 31 & October 31 & October 31 \\ 
    \hline 
    Time at 1~au (UT) & 11:14:00 & 02:23:40 & 19:43:00 & 10:48:40 & 12:08:40 \\ 
    \hline
    \end{tabularx} 
    \label{table:Parameters}
\end{table*}

\subsection{Event 2} \label{event2} 
It is also possible that performing the CME fitting at higher altitudes than for Event~1 would improve our modelling results even more. In order to test this hypothesis, we studied Event~2, for which the WL observations were available for the GCS fitting at even larger heights than 7.8~R$_{\odot}$ (Sect.~\ref{event2}).
The second event that we studied was the CME-flare observed on October 28 2021. Different aspects of this event were studied by following authors \cite{Klein22}, \cite{Guo23}, and \cite{Chikuva23}. 

\subsubsection{CME-flare on October 28 2021} \label{obs2} 

At the time of Event~2, six numbered ARs were detected on the visible side of the Sun, observed from Earth. The largest and most complex AR was the source region of the studied CME-flare event. The X1.0 GOES flare (Fig.~\ref{fig:GOESflare}a) started at around 15:20~UT, had its peak approximately at 15:30~UT, and lasted, similarly to Event~1, for several hours. Panels b and c in Figure~\ref{fig:GOESflare} show the flare observed by SDO/AIA at 304, 193, and 211~\AA. The flare originated from the NOAA AR 2887 (S26W07), which at the time of the event had a complex $\beta\gamma$ photospheric magnetic field configuration (Figs.~\ref{fig:GOESflare}b and \ref{fig:20211028_neutralline}). Several C- and M-class flares originated from the same AR earlier on the same day. The studied CME-flare event was also associated with a particle event that lasted around three days \citep[discussed in publication by][]{Klein22}. Only two ARs were situated nearby the source region of the studied flare — NOAA ARs 12888 (S12E15) and 12889 (S24E20) — and both of them had rather simple photospheric magnetic field configurations, $\alpha$ and $\beta$, respectively. No flares originating from those two ARs were associated with Event~2. Similarly to Event~1, in this event the global solar magnetic field configuration was also rather simple. The ambient magnetic field configuration, which influences the low coronal propagation of the studied CME, was influenced only by those two neighbouring ARs.

\begin{figure*}[htb]
    \centering
    \includegraphics[width=0.25\linewidth]{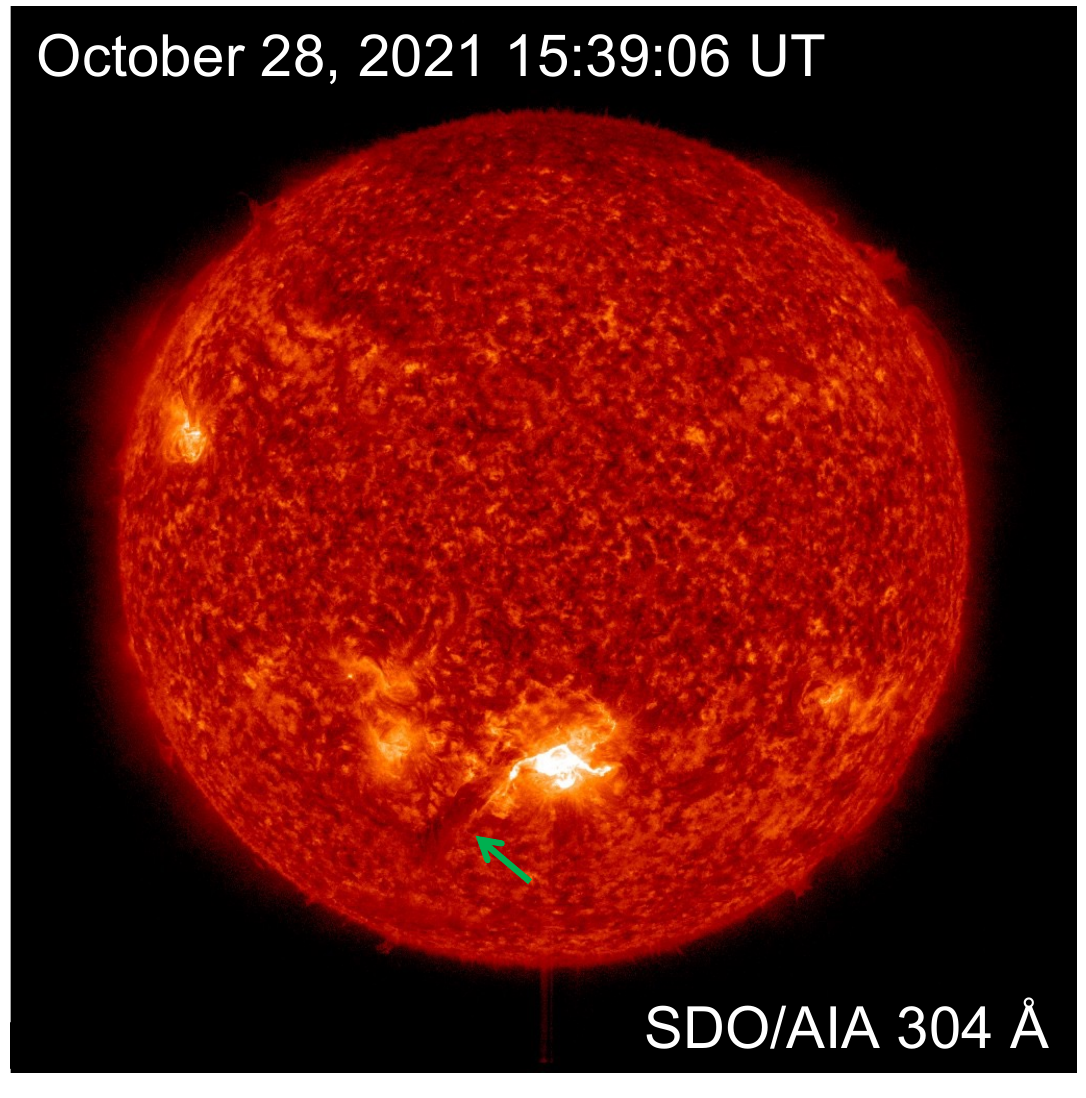}\includegraphics[width=0.25\linewidth]{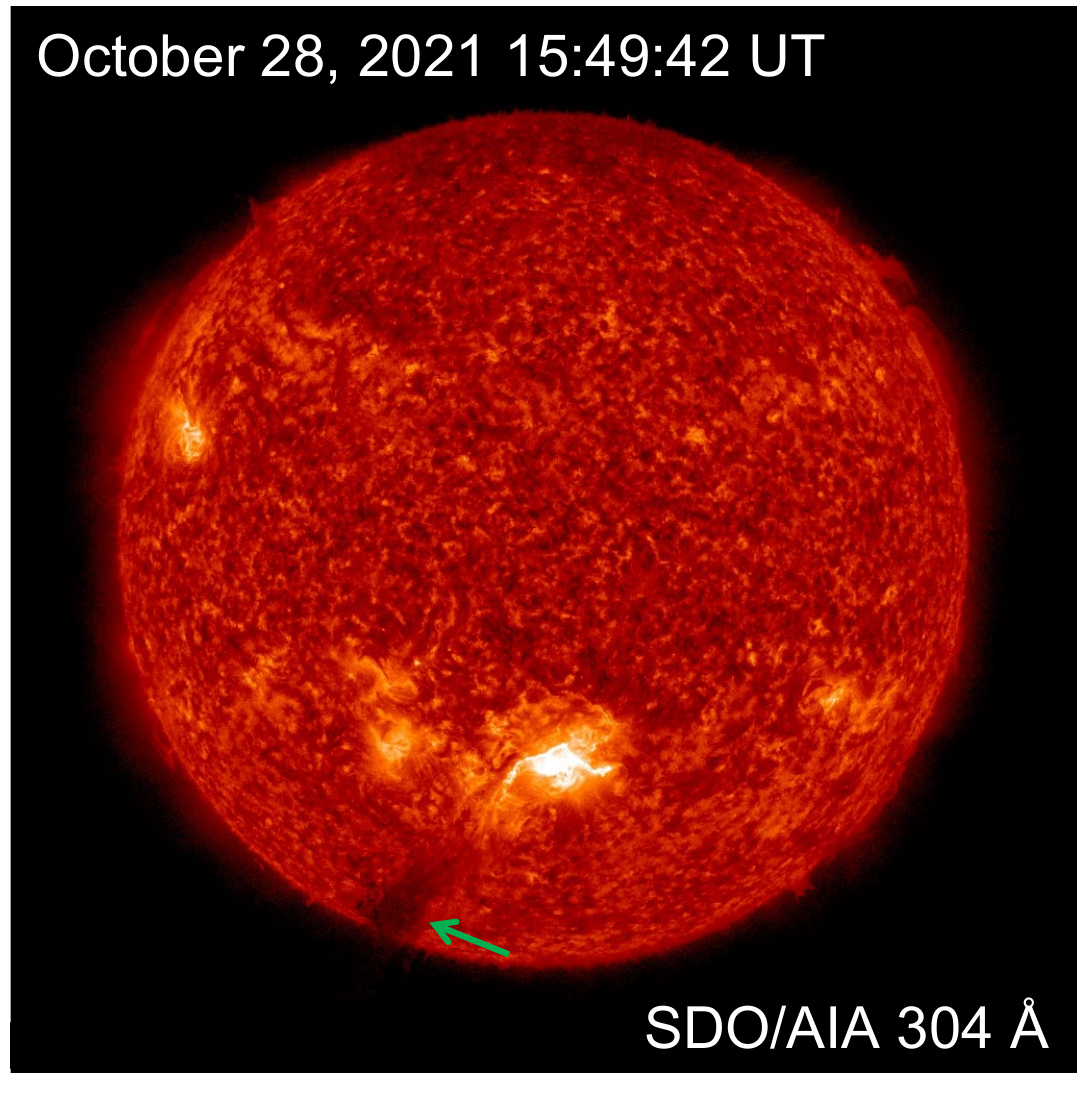}\includegraphics[width=0.25\linewidth]{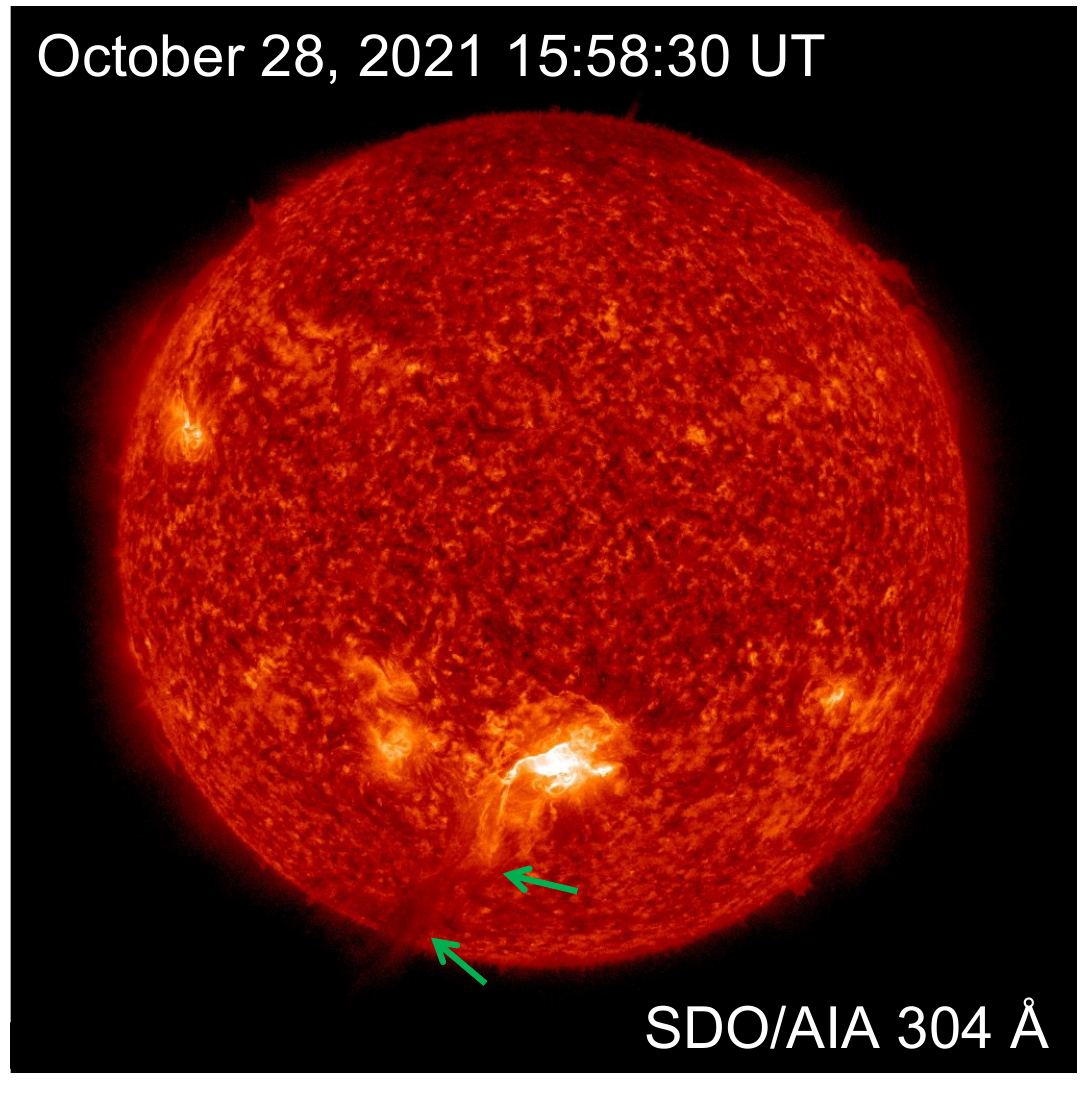}\includegraphics[width=0.25\linewidth]{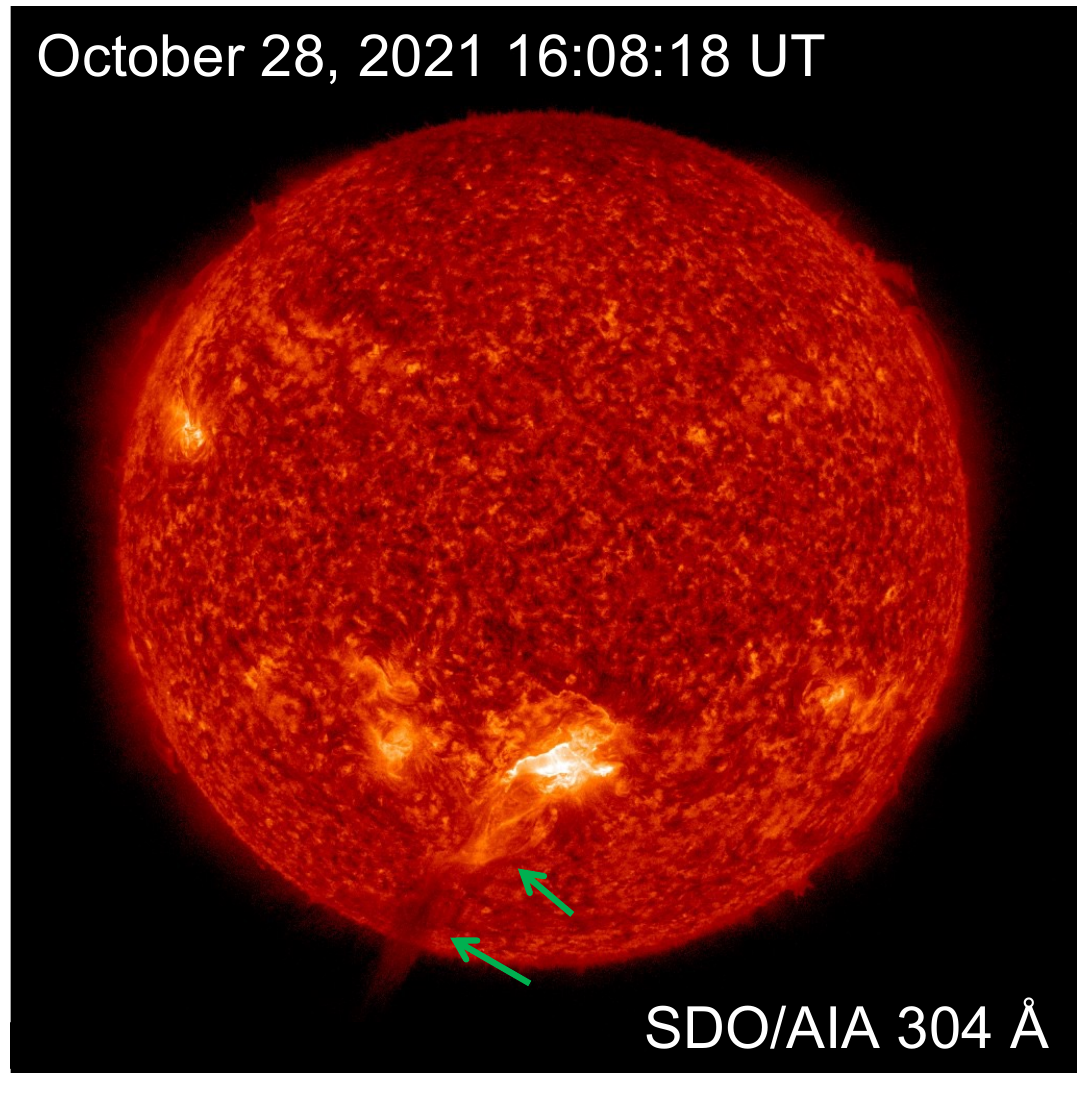}
    \caption{EUV observations taken by SDO/AIA in the 304~\AA\,channel. The green arrows indicate the position of the erupting filament during the early stage of the eruption, in which its strong southward propagation direction is clearly observed.}
    \label{fig:South_Propagation}
\end{figure*}

The studied CME was first seen in SOHO/LASCO~C2 field of view at 15:48~UT (Fig.~\ref{fig:GOESflare}c) and in STEREO~A/COR2 field of view at around 15:53~UT. The eruption of the associated filament (marked with green arrows in Figs.~\ref{fig:GOESflare} and \ref{fig:South_Propagation}) was observed by the SDO/AIA, as well as the EIT wave (marked with white arrows in Fig.~\ref{fig:GOESflare}) and the coronal dimming, the on-disc signatures of the CME. The CME-driven shock wave was observed in the WL coronagraph images by both STEREO~A and SOHO/LASCO (Fig.~\ref{fig:GOESflare}c). The narrow CME observed above the northwest solar limb is a separate event that originated on the far side of the Sun, and it therefore does not directly affect the propagation of Event~2. 

Figure~\ref{fig:South_Propagation} shows the filament eruption, marked with the green arrow at 304~\AA\,, observed by the SDO/AIA. These observations allow one to better estimate the main direction of propagation during the early stage of the eruption. They show that the CME had a well-pronounced eastward component due to the eastward orientation of the ejected filament. The arrival of the CME-driven shock wave at Earth was detected in the in situ observations at approximately 09:30~UT on October 31 2021. All the plasma parameters of Event~2 show the evidence of the FF shock wave passage (Fig.~\ref{fig:Celias}), and they fully comply with the shock detection criteria used in the heliospheric shock waves database (\citealt{Kilpua15}). If we employ equations (1) to (4), we obtain for the shock the parameters V$_{sh}$=23.9~km~s$^{-1}$, N$_{sh}$=2.63, T$_{sh}$=2.32, and B$_{sh}$=2.05, and we see that they all comply with the criteria for being considered a shock. The \href{https://space.umd.edu/pm/figs/figs.html}{SOHO/Celias} database reports this shock with a confidence level of 99~\%. 

The radio observations of Event~2 show a complex metric type~II burst and a short-duration but intense interplanetary type~II radio burst (Fig.~\ref{fig:swaves}b). 
The type~II burst was preceded by very intense type~III radio bursts observed by both STEREO/WAVES and WIND/WAVES instruments. The shock-associated emission started at 15:37~UT and lasted until around 15:57~UT. A low-frequency type~II burst was also observed, similar to the one reported by \cite{Jebaraj20}. 

\subsubsection{CME fitting and modelling with EUHFORIA} \label{model2} 

The input parameters for the EUHFORIA model were obtained by employing the GCS fitting method (\citealt{Thernisien06}; \citealt{Thernisien09}; \citealt{Thernisien11}). As for Event~1, we also checked the orientation of the AR neutral line in the SDO/HMI magnetogram (Fig.~\ref{fig:20211028_neutralline}c), the orientation of the filament before eruption, and the orientation of the post-flare loops (SDO/AIA observations at 304 and 193~\AA), in order to perform the GCS fitting as accurately as possible. The orientation of the main neutral line, which shows a somewhat different orientation than the filament, is marked at 304~\AA\,at 15:31:53~UT, at 193~\AA\, at 15:32:16~UT, and in the magnetogram at 15:31:57~UT (Fig.~\ref{fig:20211028_neutralline}). This procedure allowed us to employ the correct orientation of the CME legs during the GCS fitting, which was in this event more similar to the filament orientation than to the main neutral line of the AR. We performed five GCS fittings (GCS1 to GCS5) using pairs of STEREO/COR2 and SOHO/LASCO~C2 images at almost simultaneous times. These five pairs of images covered the full radial distance observed by the coronagraphs. The example of the GCS fitting is presented in the Appendix. (Fig.~\ref{fig:Example_fit_Event2}). Table~\ref{table:FittingsOctober} presents the values of the obtained CME parameters for different pairs of images. The speed of the CME was calculated in the same way as for Event~1 (Sect.~\ref{model1}) and it amounts to 1\,130, 1\,350, 1\,080, 1\,010 and 1\,010 km~s$^{-1}$ for GCS1 to GCS5, respectively. These speeds were used to obtain the time of the CME at 0.1~au, the insertion height of the CME in EUHFORIA model. The kinematic properties of Event~2 using different height and time points is presented in Fig.~\ref{fig:swaves}d. The symbols and the colours of the data points and of the curves represent the same quantities as in Fig.~\ref{fig:swaves}c (see Sect.~\ref{obs1}). For the modelling of Event~2, together with the GCS fitting, we also employed CME parameters obtained from the DONKI database. The input parameters from the DONKI database and the two last GCS fittings, GCS4 and GCS5, are shown in Table~\ref{table:Parameters}. 

\begin{figure*}[htb]
    \centering
    \includegraphics[width=1.0\linewidth]{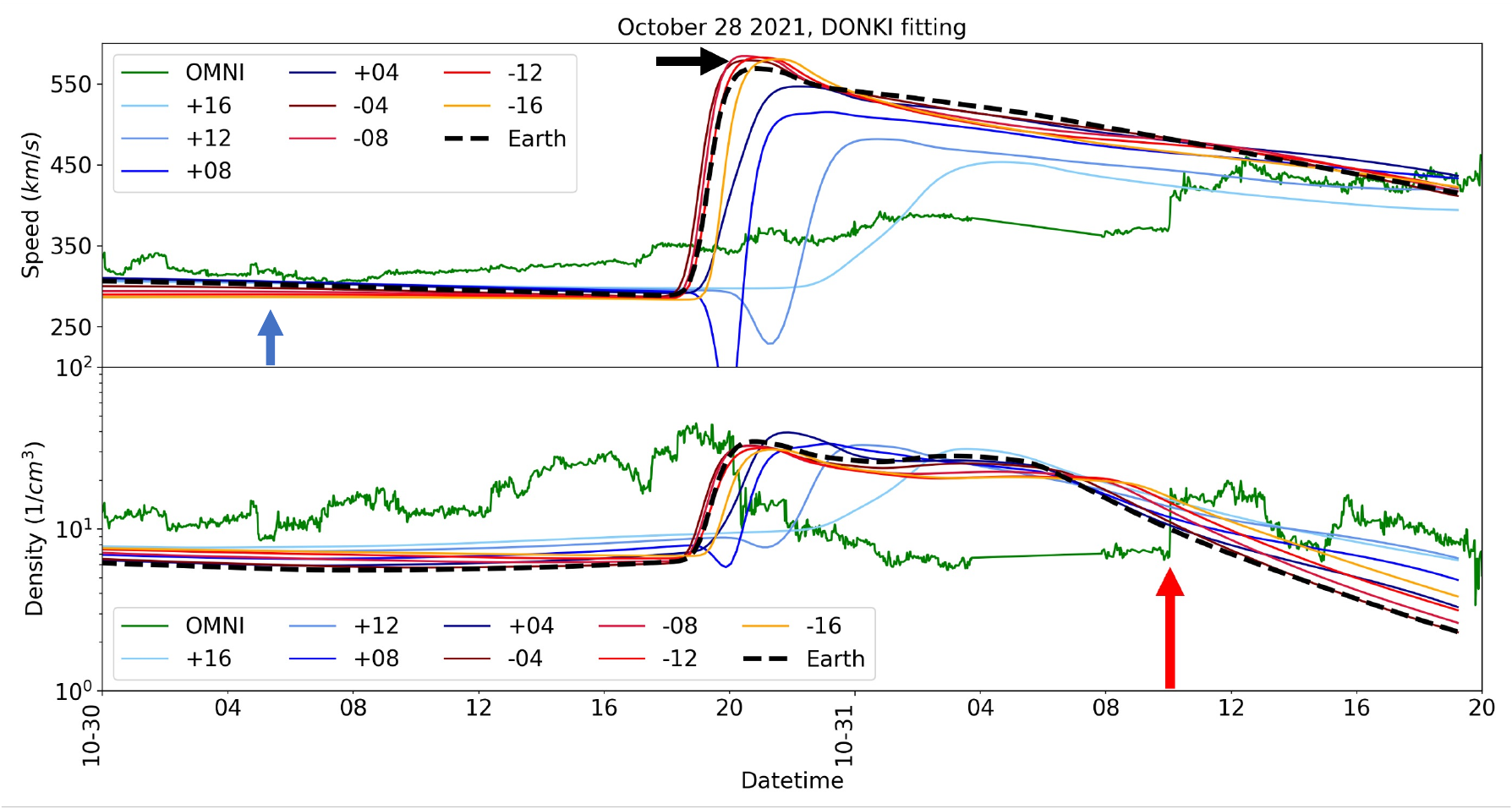}
    \caption{Comparison of in situ observations at Earth and modelled velocity (top panel) and density time series (bottom panel) obtained in EUHFORIA simulations for Event~2. The in situ data are presented with the green line and the modelled time series have different shades according to the positions at which they were estimated (see also Fig.~\ref{fig:STEREO_pos}a). The modelling results were obtained using as inputs the CME characteristics obtained from the DONKI database. The red arrow indicates the observed arrival time at Earth. The black arrow indicates the modelled arrival time of the shock wave at Earth (dashed black curve) at zero degree latitude and longitude. The blue arrow indicates the estimated arrival time, employing the drift rate of the type~II radio bursts. The modelled arrival time of the shock wave at Earth is significantly earlier than the observed one, $\Delta$t~$\approx$~-12~h. The observed $\Delta$t is different for different latitudes, ranging from 7--12~h.}
    \label{fig:Donki_Orig}
\end{figure*}

For the optimisation of the solar wind modelling (see Sect.~\ref{methods}), we used three GONG synoptic magnetogram maps of October~26 2021 at 18:00 and 23:24~UT and of October~27 at 05:24, 13:04, and 19:04~UT. The aim of this procedure was to find the magnetogram that would facilitate accurate modelling results, in good agreement with the observed in situ solar wind. The selected magnetogram was at 19:04~UT on October~27. Following the optimization of the solar wind modelling with EUHFORIA, we modelled the CME. The in situ time series of the solar wind plasma parameters at Earth, together with the results of the CME modelling with the DONKI parameters, are presented in Fig.~\ref{fig:Donki_Orig}. The panels are in the same order, and the curves and the arrows represent the same quantities as in Fig.~\ref{fig:December_DONKI}. Similar to the findings of Event~1 (see Sect.~\ref{model1}), for Event~2 we also find that, in the case of the CME parameters from the DONKI database, the modelled shock arrival is significantly earlier than the observed one (indicated in Fig.~\ref{fig:Donki_Orig} with black and red arrows, respectively). The difference in this case is $\Delta$t~$\approx$~-12~h, while the amplitude of the modelled shock was about three times bigger than in observations in the case of the speed, and about one and a half times bigger for the density. Those results are an indication that the modelling of Event~2 is not very accurate when the CME parameters are obtained from the DONKI database. Seemingly, the CME speed provided in the DONKI database is higher than the actual CME speed. Additionally, differences with other CME parameters obtained from the DONKI database and the real CME parameters can be expected. 

The different latitudes in the vicinity of Earth, above (blue shades) and below Earth (red shades), also show a difference in $\Delta$t, with a range of inaccuracy of about 6-14~h. For Event~2, as for Event~1, the latitudes that are the furthest from Earth ($\pm$16$^{\circ}$), which are the locations of the impact of the flanks of the CME, have modelled arrival times closer to the observed ones (Fig.~\ref{fig:Donki_Orig}). Also, Event~2, like Event~1, has a main propagation direction strongly southward from the Sun-Earth line. This is also indicated by the characteristic of the blue curve at +16$^{\circ}$, which shows an almost flat profile compared to the other blue and red curves. 

After the modelling with the CME input parameters from the DONKI database, we input the sets of the CME parameters obtained in GCS1 to GCS5 (Table~\ref{table:FittingsOctober}) to EUHFORIA with the same background solar wind. The changes in the values of the different fittings for the same parameters indicate, similarly to Event~1, that the main propagation direction of the CME changes as the CME propagates away from the Sun. This change affects the estimation of the CME speed as well as the time at which the CME arrives at Earth.  
 
\begin{table*}
    \caption{CME parameters obtained from the GCS fittings, necessary as an input for the Event~2 CME simulations. The time of observation 1 corresponds to the time of CME observations in the LASCO~C2/C3 field of view, and the time of observation 2 to the CME observed by STEREO~A/COR2.} 
    \centering
    \begin{tabularx}{0.95\linewidth}{@{\extracolsep{\fill}}|l|r|r|r|r|r|r|r|} 
    \hline
    \multicolumn{8}{|c|}{October 28, 2021} \\
    \hline
    Parameters & GCS1 & GCS2 & GCS3 & GCS4 & GCS5 & Mean & Stdev \\ 
    \hline\hline
    Time in LASCO (UT)& 15:48:07 & 16:24:08 & 16:36:07 & 16:54:07 & 17:18:07 & - & - \\
    \hline 
    Time in STA (UT)& 15:53:30 & 16:23:30 & 16:38:30 & 16:53:30 & 17:23:30 & - & - \\
    \hline 
    Time at 0.1\,AU (UT) & 19:06:47 & 18:19:00 & 18:48:00 & 18:58:00 & 18:45:00 & - & - \\ 
    \hline
    Latitude ($^{\circ}$) & -25 & -29 & -39 & -42 & -44 & 35.6 & 8.6 \\ 
    \hline
    Longitude ($^{\circ}$)  & 0 & 0 & 0 & 0 & 0 & 0.0 & 0.0 \\ 
    \hline
    Height ($R_{\odot}$)    & 3.5 & 7.0 & 8.4 & 9.7 & 12.3 & - & - \\
    \hline 
    Half Angle ($^{\circ}$) & 49 & 49 & 49 & 49 & 49 & 49.0 & 0.0 \\
    \hline 
    Tilt                    & 20.8 & 31.3 & 45.1 & 45.7 & 44.6 & 37.5 & 11.1 \\
    \hline 
    Speed (km~s$^{-1}$)            & 1130 & 1350 & 1080 & 1010 & 1010 & - & - \\ 
    \hline 
    Day at 1 AU & October 30 & October 30 & October 31 & October 31 & October 31 & - & - \\ 
    \hline 
    Time at 1 AU (UT) & 21:43:40 & 16:23:40 & 06:08:40 & 10:48:40 & 12:08:40 & - & - \\ 
    \hline
    \end{tabularx} 
    \label{table:FittingsOctober}
\end{table*}

For Event~2, the standard deviation for the latitude amounts to a significant 8.6~\%, while for the longitude is zero. This shows that the deflections experienced by the CME change its main direction of propagation towards the south from the Sun-Earth line. The CME half-angle does not change, which means that the CME does not significantly expand at the considered heights. The tilt also strongly changes (by about 11~\%), which indicates that the CME also rotates and changes its direction of propagation as it evolves. The changes in this case are larger than for Event~1, and consequently they affect the direction of propagation of the CME more. The southward direction of propagation of the CME is clearly visible in the EUV data also (see Fig.~\ref{fig:South_Propagation}), but it was difficult to quantify it and it does not influence the estimation of the CME parameters using the GCS fitting, which starts at larger heights. 

\begin{figure*}[htb]
    \centering
    \includegraphics[width=1.0\linewidth]{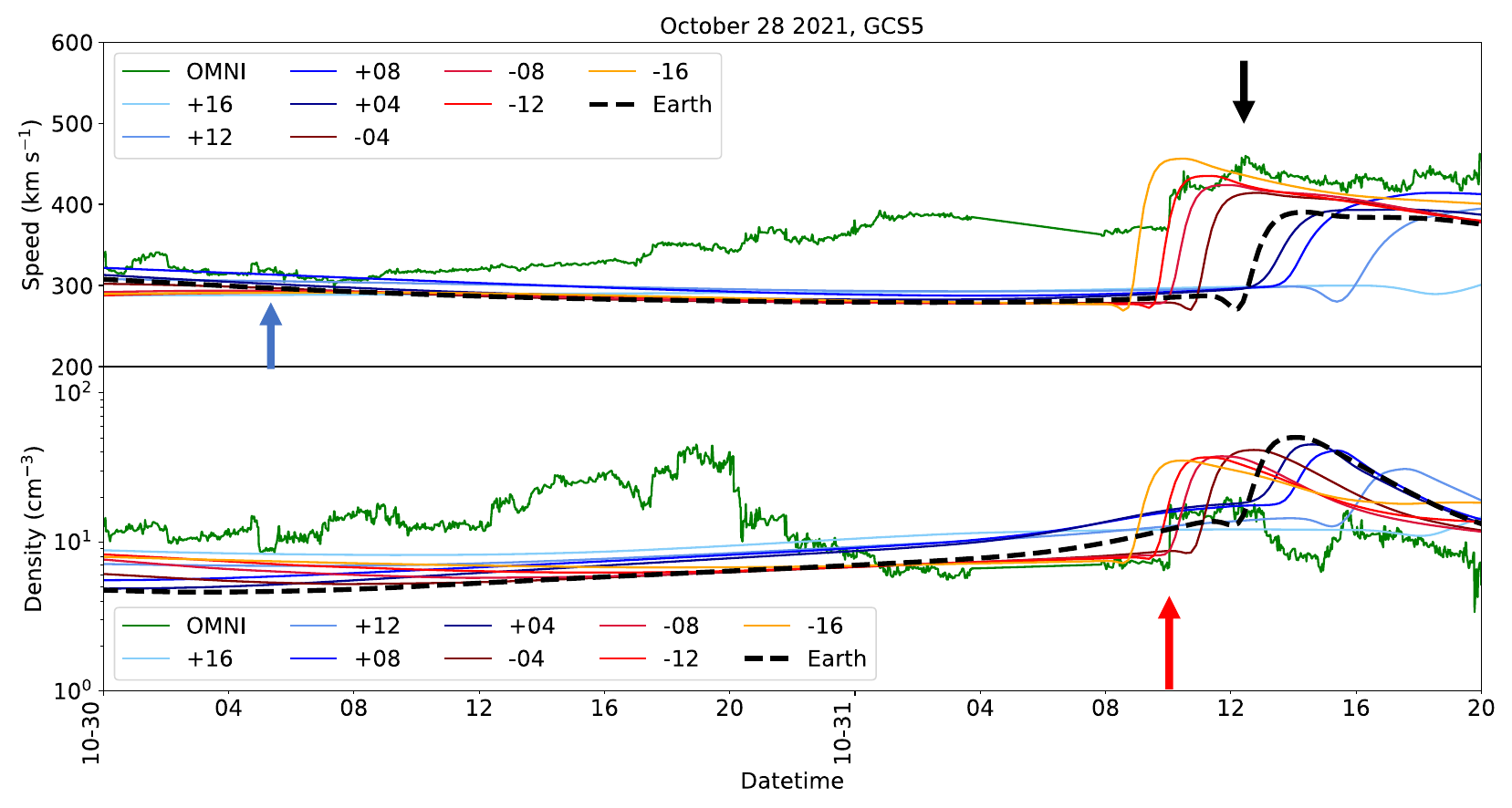}
    \caption{Comparison of in situ observations at Earth and modelled velocity (top panel) and density time series (bottom panel) obtained in EUHFORIA simulations for Event~2. The in situ data are presented as the green line and the modelled time series have different shades according to the positions at which they were estimated (see also Fig.~\ref{fig:STEREO_pos}a). The modelling results were obtained using the CME characteristics obtained with GCS5. The red and black arrows mark the observed arrival time at Earth and the modelled arrival time of the CME (also a dashed black curve) at zero degree latitude and longitude, respectively. The blue arrow indicates the estimated shock arrival time obtained from the drift rate of type~II radio burst. The modelled arrival time of the shock wave at Earth is close to the observed one, $\Delta$t~$\approx$~+1~h. The observed $\Delta$t is different for different latitudes, ranging up to 15~h.}
    \label{fig:GCS-44}
\end{figure*}

Figure~\ref{fig:GCS-44} shows the results of the simulation that employed CME parameters obtained in the last feasible GCS fitting; that is, GCS5 at the greatest heights of the STEREO/COR2 field of view. The time series at the positions below Earth (red shades) show earlier arrival time of the CME event than the time series above Earth (blue shades). Additionally, the profile of the disturbance is much steeper in the case of below-Earth time series than for the ones above Earth, indicating that the major part of the CME passes below Earth. Namely, the main propagation direction of the studied CME is south of the Sun-Earth line. 

\begin{figure*}[htb]
    \centering
    \includegraphics[width=1.0\linewidth]{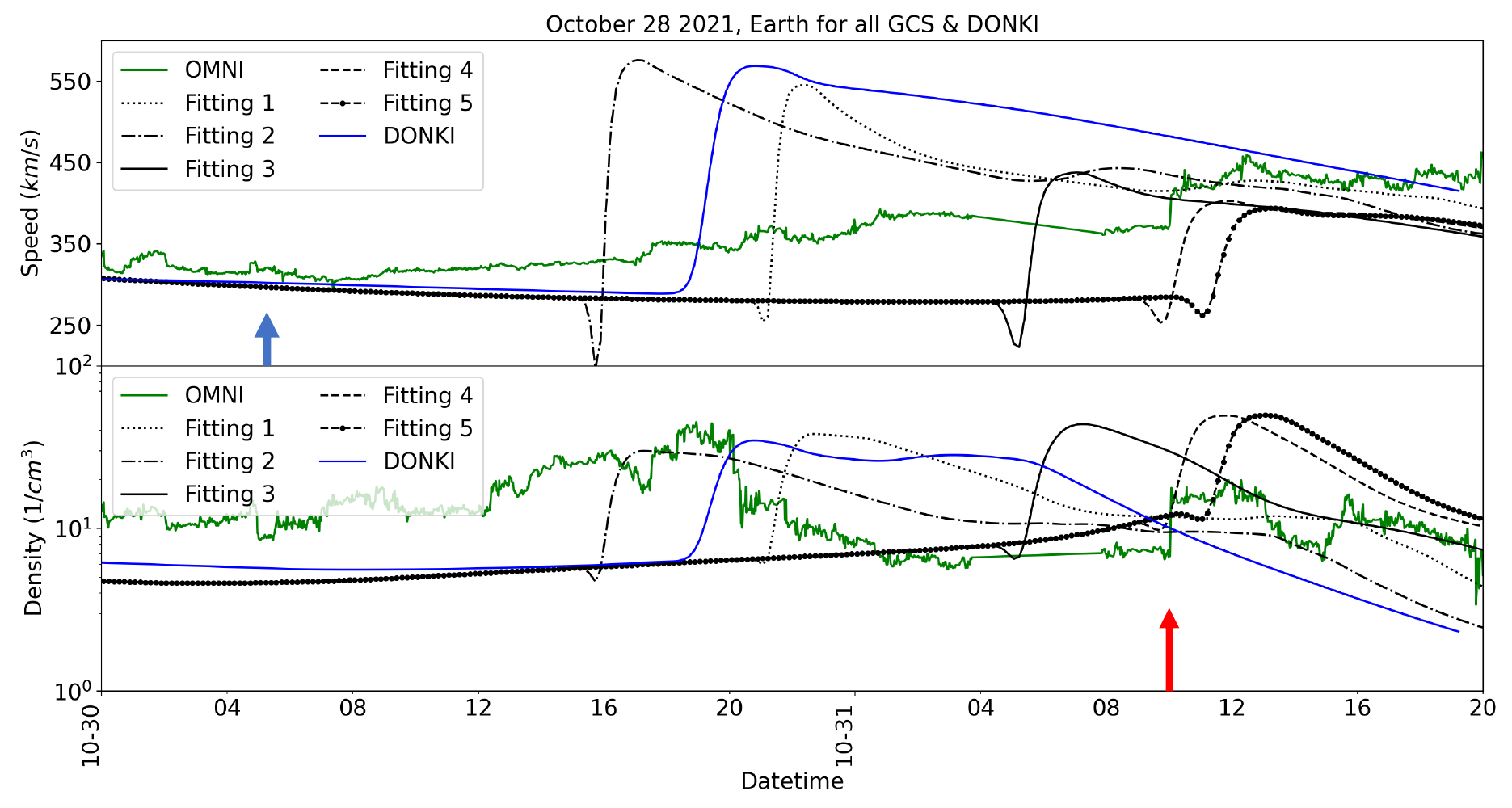}
    \caption{Comparison of in situ observations at Earth and modelled velocity (top panel) and density time series (bottom panel) obtained in EUHFORIA simulations, for the cases of the Earth, at zero degree longitude and latitude, for Event~2. The in situ data are presented as the green line and the modelled time series have different line styles according to the fitting that they present. We also include with the blue curve the results for Earth, for the case of the DONKI parameters. The modelling results were obtained using the CME characteristics obtained from the fittings that we obtained with the GCS technique. The red arrow indicates the observed arrival time at Earth. The blue arrow indicates the estimated arrival time from the measurements of the radio observations. The modelled arrival times of the shock waves at Earth are at earlier times at first but they gradually shift towards the observed one as the fittings are done for times that the CME is further from the solar surface. The observed $\Delta$t is different for different fittings, ranging from 2-18~h.}
    \label{fig:earth_all}
\end{figure*}

Figure~\ref{fig:earth_all} is the same as Fig.~\ref{fig:December_earth_all}, but for Event~2, showing the modelled arrival time of the shock at Earth for five different parameter sets (Table~\ref{table:FittingsOctober}). The modelling results of GCS4 and GCS5 are quite similar with a bit more than 1~h of difference in the time of arrival at Earth when comparing them with each other. The arrival time of the CME obtained with the GCS4 parameters is almost simultaneous with the observed shock arrival (Fig.~\ref{fig:earth_all}), while GCS5 provides results with a bit more than 1~h of delay. This result can be induced by a few factors acting simultaneously. At the heights of 9.7--12.3~R$_{\odot}$ at which GCS4 and GCS5 were performed, the major part of the rotation, deflections, and distortions that can influence the propagation of the CME is expected to be finished. Therefore, we do not expect a significant change in the CME parameters estimated at close to the edge of the field of view of the SOHO/LASCO coronagraph. Additionally, at these heights the CME structure starts to be more diffuse and it is often hard to clearly distinguish the CME leading edge, which induces additional uncertainties in the GCS fitting, whose results are anyway very strongly dependent on the person performing the reconstruction (see e.g. \citealt{Verbeke23}). The difference between the observed and modelled arrival time for the fitting GCS5 is $\Delta$t~$\approx$~1~h, which is still a very good modelling accuracy. In order to understand what the heights are after which the CME does not significantly change its propagation direction, and how exactly the change in the different CME parameters influences the accuracy of the modelled arrival time, a statistical analysis will be performed. We note that, like in Fig.~\ref{fig:Donki_Orig}, the dips that appear in the plot of the speed are the numerical artefacts with no physical meaning. 

For Event~2, the only parameters that significantly change in the GCS fittings are the latitude of the source region of the CME and the CME's tilt. Figure~\ref{fig:earth_all} shows a significant improvement in the modelled arrival time of the shock when the fittings are done at increasing heights. The most accurate forecast is obtained with GCS4 input parameters, as was discussed earlier. A systematic decrease in the amplitude of the modelled CME arrival (top panel of Fig.~\ref{fig:earth_all}) is observed when we employ the CME input parameters obtained at larger heights. This result arises because the angle between the Sun-Earth line and the main propagation direction of the CME is increasing. Accordingly, the part of the modelled CME that is observed near Earth is deviating more and more from the CME's leading edge, towards its north flank. 

\subsubsection{RADIO observations}
The radio event associated with the CME on December 28 2021, briefly described in Sect.~\ref{obs2}, consisted of intense type~II and type~III radio bursts, both starting already in the metric range; that is, observed by ground-based instruments. In contrast with Event~1, for Event~2 the type~III bursts (at around 15:30~UT) associated with the flare impulsive phase were more intense. The type~II burst started at about 15:37~UT and lasted for about 20 minutes (Fig.~\ref{fig:swaves}b). The ending of type~II was at about 6~MHz. The low-frequency type~II burst was also observed starting at about 30~MHz, similar to the one reported by \citealt{Jebaraj20}. 

In order to obtain the velocity of the CME-driven shock, we employed the type~II frequency drift rate and the coronal electron density model, like for Event~1 (see Sect.~\ref{radio1}). The estimated velocity from the radio observation was found to be about 1\,040~km~s$^{-1}$, which is similar to the CME velocity estimated using a GCS fitting (Table~\ref{table:FittingsOctober}). This type~II velocity gives the arrival time of the associated shock at Earth at around 4:55~UT on October 30. We indicate this time in Figs.~\ref{fig:Donki_Orig}, \ref{fig:GCS-44} and \ref{fig:earth_all} with a blue arrow, like for Event~1. Comparing this arrival time with the in situ time series (Fig.~\ref{fig:GCS-44}), we see that the radio observations give the early arrival time of the shock, $\Delta t\approx-29$h. This large error in the estimated arrival time, together with the similarity between the CME and type~II velocity, indicates the possibility that the type~II emission is generated near the leading edge of the CME. On the other hand, due to the short duration of the type~II burst and its complex multi-lane structure, it is possible that the obtained type~II velocity is overestimated.
Positioning the type~II radio sources in 3D space, which is out of the scope of this study, is necessary to verify the reason behind such a large difference in the shock arrival time (see e.g. \citealt{Magdalenic14}, \citealt{Jebaraj20}).

The 2D speed for Event~1 that is given in the \href{https://cdaw.gsfc.nasa.gov/CME_list/index.html}{SOHO/LASCO} catalogue is 1\,519~km~s$^{-1}$ at a height of around 23~R$_\odot$. This results in an arrival time for the shock at Earth on October 29 at around 18:42~UT. 

\section{Summary and conclusions} \label{Results}
The accurate modelling of heliospheric conditions for forecasting purposes is one of the important topics in space weather science. The main purpose of different models developed in recent decades is accurate forecasting of the arrival of CME, CME-driven shock, and solar energetic particles at Earth, and the effects induced by their interaction with Earth's magnetosphere. 

In this study, we aim to improve the accuracy of the modelled time of CME arrival at Earth employing the EUHFORIA and the Cone-CME model. The majority of existing CME and solar wind models, including EUHFORIA, launch the CME at the height of 0.1~au. They employ CME parameters obtained at distances close to the Sun, considering predominantly the radial direction of the CME's propagation at heights below 0.1~au. As a large fraction of CMEs have in the low corona a main propagation direction that is strongly different from the Sun-Earth line, such an approximation strongly influences the accuracy of forecasting results. 

In order to understand how important the main propagation direction of the CME is when estimating the arrival of the CME and shock wave at Earth, we employed three very different methods. The first two methods are rather simple and straightforward. They are compared with sophisticated and rather realistic modelling of CMEs.

The first method employs estimation of the CME's arrival using the 2D velocity of the CME (\href{https://cdaw.gsfc.nasa.gov/CME_list/index.html}{SOHO/LASCO}) obtained from WL observations. 
In the second method, we used the drift rate of type~II bursts, signatures of the CME-driven shock waves, and the 1fold \cite{Leblanc98} coronal electron density model to estimate the speed of the associated shock wave and accordingly the arrival of the shock at 1~au. 

The third method, which is the main focus of our study, is more sophisticated and includes modelling of the CMEs with EUHFORIA and the CME cone model. In this method, we considered a few sets of the CME input parameters obtained in two different ways:

a) The DONKI database, which is an open access catalogue of CMEs, providing the source region of the CME (longitude and latitude), its speed, its half-width, and the time of the CME at 0.1~au. 

b) Fitting with the GCS technique, which is a rather accurate method (in the case of observations from widely separated spacecraft) to obtain the CME parameters needed for the cone model. The advantage of this technique lies in the fact that it uses coronagraph observations to help fit a mesh~grid on the observed body of the CME. The coronagraph observations show the CME in the low corona, which are the heights at which the CME still experiences deflections, rotations, and other distortions conditioned with the ambient low coronal structures, which can substantially affect a CME's direction of propagation. By fitting coronagraph observations at different times, and hence at different heights, the change in the direction of propagation can be mapped and taken into account when modelling CMEs. One of the important CME parameters needed in the fitting with the GCS is also the tilt, which indicates the orientation of the CME legs. In order to accurately retrieve an initial tilt value for the first GCS fitting, we used magnetograms of the AR, which sourced the CME to find the position of the neutral line (see Sects.~\ref{model1} and \ref{model2}). Table~\ref{table:Parameters} lists the CME parameters used in the two most important runs for each studied event. 

We present the multiwavelength study and modelling with EUHFORIA, of the CME-flare events that were observed on December 7 2020 and October 28 2021, with the aim of taking into consideration the contribution of the change in the CME's direction of propagation in the low corona; that is, below 0.1~au.  
The EUV and WL observations of the corona at the time of both studied events show flaring, prominence eruptions, associated waves, dimmings, and halo CMEs, with the bulk of the CME mass propagating southwards from the Sun-Earth line. In both events, a weak shock wave associated with the studied CME was observed in the in situ data at 1~au. The weak shock signatures are an indication of the possibility that only the CME flanks impacted Earth because the CMEs did not propagate along the Sun-Earth line but deviated from it to some degree. 

The accuracy of the estimated arrival time of the CME and the shock at 1~au is summarised in Table~\ref{table:TimeDifferences}. We also list and discuss our results in more detail below:

1) The largest inaccuracy was found when employing the 2D velocity (\href{https://cdaw.gsfc.nasa.gov/CME_list/index.html}{SOHO/LASCO} catalogue) obtained from the WL images with $\Delta$t amounting to $-29$h and $-39$h, in the cases of Event~1 and Event~2, respectively. We note that this is one of the most frequently employed methods of estimating the CME's arrival at Earth. However, when only one point of view is used, the estimation of the CME velocity is subject to significant errors due to projection effects. This is particularly important in the case of the halo, or partial halo CMEs, which are the most important in space weather forecasting. We can conclude that the use of a pair of almost simultaneous images, from two different viewing points, is very important for the accurate estimation of the timing of a CME's arrival at Earth.

2) The radio observations provided somewhat better results, but for Event~1 the estimated shock arrival was too late ($\Delta$t~$\approx$~+21~h), and for Event~2 it was too early ($\Delta$t~$\approx$~-29~h) when compared with the in situ observations. This inconsistency in the estimated shock arrival depends on the accuracy of the estimation of the drift rate and on the relative position of the source of the radio emission and the CME; that is, CME flanks versus the leading edge of the CME (e.g. \citealt{Magdalenic14}). 

3) All of the modelling results provide a better estimation of the arrival time of the CME at 1~au than the 2D CME velocity and the type~II radio bursts (Table~\ref{table:TimeDifferences}). 

The modelled arrival time when using the CME input provided in the DONKI database was earlier than the observed arrival time in the in situ data, with $\Delta$t amounting to about $-14$h and $-12$h for Event~1 and Event~2, respectively. Modelling of the CMEs employing the CME input parameters obtained in different GCS fittings at increasing heights shows a constant improvement in the accuracy of the predicted arrival time, as the CME is propagating away from the Sun. The best modelling results, with only 10 to 30 minutes of inaccuracy, were obtained when the CME fitting was done close to the edge of the field of view of the coronograph (at a height of about 8-10~R$_{\odot}$). 

In the case of Event~1, the changes in the CME parameters, which were obtained by fitting at larger heights subsequently, mostly affect the latitude and longitude of the CME. These two parameters relate to the position that the source region of the CME would have if the CME were to propagate predominantly radially. We also found that the change in the half-width of the cone corresponds to the CME expansion as it evolves. The speed of Event~1 increases in consecutive fittings, reaching a final value of 1\,240~km~s$^{-1}$, indicating that the CME is still accelerating in the low corona.

In the case of Event~2, the change in the CME parameters at the larger heights is almost entirely in the latitude, indicating that the main propagation direction of the CME is increasingly southward as the CME propagates to larger heights. This behaviour is probably influenced by the neighbouring coronal structures that influence the low coronal propagation of the CME. The CME speed for Event~2 increases from GCS1 to GCS2 and then slowly decelerates to a value of 1\,010~km~s$^{-1}$ (GCS4 and GCS5), indicating a shorter CME acceleration phase than for Event~1. 

We summarise the most important findings of our study related to the CME modelling with EUHFORIA and the cone CME model:

\begin{itemize}
    \item [\textbullet] 
When we used the CME input parameters obtained in the subsequent GCS fittings and at the consecutively larger heights, the accuracy of the modelled arrival time drastically improved (Table~\ref{table:TimeDifferences}). We found $\Delta$t changing from 20~h and 12~h to as little as 10~min and 30~min, for Event~1 and Event~2, respectively. 
We note that for Event~2 the best result was obtained for the semi-last fitting, GCS4, and the accuracy worsened when the last possible fitting was employed. Possible reasons for such behaviour are discussed in Sect.~\ref{model2}.
We can conclude that the GCS fitting that we did when the CME was further away from the solar surface (Table~\ref{table:TimeDifferences}) provides the CME parameters that yield very accurate modelling results with EUHFORIA. The main reason for this is that the CME deflections and rotations, which affect its direction of propagation, are due to its interaction with ambient low coronal structures (e.g. other ARs, coronal holes, or helmet streamers). Those interactions happen low in the corona, so when for the fitting later observations are used, the CME is at or close to its final direction of propagation. This shows the importance of reconstructing the CME at larger heights, preferably at around 10~R$_{\odot}$ (Table~\ref{table:TimeDifferences}), and the estimation of the correct propagation direction of the CME.
\end{itemize}

\begin{itemize}
    \item [\textbullet]The modelled arrival time when using the CME input provided in the DONKI database was for both events earlier than the observed arrival time in the in situ data, with $\Delta$t~<~$-10$~h. This indicates that the DONKI parameters can be employed for the CME modelling and estimation of the arrival time to Earth, but with limited accuracy.
\end{itemize}

\begin{itemize}
    \item [\textbullet]Estimating the correct tilt of the CME by inspecting the orientation of the neutral line in the source region is very important when performing GCS fitting. Constraining the CME tilt helps to properly orient the flux rope and then fit the other parameters (the longitude, latitude, height, and half-angle) depending on that. Both studied events show a significant change in the tilt value, which is indication of the rotation that also affects the CME's direction of propagation. This result shows that although the initial tilt value is very important, it should be not kept fixed during consecutive GCS reconstructions.
\end{itemize}

\begin{itemize}
    \item [\textbullet]The most important conclusion of our study is as follows. The estimation of the main propagation direction of the CMEs, all the way from the low corona and up to about 13~R$_{\odot}$, is very important for the accuracy of modelling the CME and CME-driven shock arrival at Earth. This consideration allowed us to drastically improve the modelled CME arrival time, up to the excellent accuracy of 10-30~min, even without using complex and more advanced CME models, but only with the simple cone CME model. Both studied events, although they were halo CMEs seen in coronagraphic observations, had a main propagation direction that significantly deviated from the Sun-Earth line. Our study stresses the great importance of correctly estimating the main direction of CME propagation for accurately forecasting the time of a CME's arrival at Earth.
\end{itemize}

In order to better quantify which of the CME parameters are more important for the accurate estimation of its main propagation direction and, further, its arrival time at Earth, it is necessary to perform statistical analysis of several events using the same method as in this study. Further, it is also very important to understand at which range of heights in the corona the CMEs have experienced the major fraction of the change in their main propagation direction, as such a height could be the reference point above which the GCS reconstruction would provide accurate information about the direction of propagation of the CME. 

\begin{table}
    \caption{Time differences in the arrival times of the CME-driven shock at Earth, for the speeds retrieved with different methods, compared to the observed arrival time.}
    \centering
    \begin{tabularx}{0.95\linewidth}{@{\extracolsep{\fill}}|l|c|c|} 
    \hline
    Parameters & Event 1 & Event 2 \\
    \hline\hline 
    Radio & +21 h & -29 h \\
     & (4.4 R$_{\odot}$) & (1.8 R$_{\odot}$) \\
    \hline 
    2D Speed & -29 h & -39 h \\ 
     (\href{https://cdaw.gsfc.nasa.gov/CME_list/index.html}{SOHO/LASCO}) & (28 R$_{\odot}$) & (23 R$_{\odot}$) \\ 
    \hline 
    DONKI & -14 h & -12 h \\
     & & \\
    \hline 
    GCS 1 & + 20 h & -12 h \\
     & (3.1 R$_{\odot}$) & (3.5 R$_{\odot}$) \\
    \hline 
    GCS 2 & +11 h & -16 h \\
     & (4.1 R$_{\odot}$) & (7.0 R$_{\odot}$) \\
    \hline 
    GCS 3 & +4 h & -4 h \\
     & (6.2 R$_{\odot}$) & (8.4 R$_{\odot}$) \\
    \hline 
    GCS 4 & + 10 min & +30 min \\
     & (7.8 R$_{\odot}$) & (9.7 R$_{\odot}$) \\
    \hline 
    GCS 5 & - & + 2 h \\
     & & (12.3 R$_{\odot}$) \\
     \hline
    \end{tabularx} 
    \label{table:TimeDifferences}
\end{table}

\begin{acknowledgements} We thank the radio monitoring service at LESIA (Observatoire de Paris) to provide value-added data that have been used for this study. The authors are thanking dr. Athanasios Kouloumvakos, the developer of the PyThea tool which is used to obtain the plots from the coronagraphs. Moreover, we would like to thank Antonio Esteban Niemela for his help, in particular with the EUHFORIA modelling, dr. prof. Stefaan Poedts for his valuable input in personal discussions and dr. Luciano Rodrigues for his interesting comments in the meetings that we had about this project. Last but not least, we thank Anna Chulaki, for the valuable information and resources that she provided us, about the way the DONKI database obtains the values for the CMEs. 
These results were obtained in the framework of the projects BR/165/A2/CCSOM, B2/191/P1/SWiM and the FED-tWIN project PERIHELION funded by the Belgian Federal Science Policy Office (BELSPO). A. V. acknowledges the financing of the PhD project nb 3E211295.
Data supplied courtesy of \href{https://www.solarmonitor.org/}{Solar monitor}. Data supplied courtesy of the SDO/HMI and SDO/AIA consortia. SDO is the first mission to be launched for NASA's Living With a Star (LWS) Program. This work utilizes magnetogram, intensity, and farside data obtained by the Global Oscillation Network Group (GONG) project, managed by the National Solar Observatory, which is operated by AURA, Inc. under a cooperative agreement with the National Science Foundation. Solar Region Summaries, Solar Event Lists, GOES 5-min X-rays, proton and electron data from the NOAA Space Weather Prediction Center. 

\end{acknowledgements}

\bibliographystyle{aa}
\bibliography{bibli} 

\begin{appendix}
\section{GCS fittings} \label{appendix}

We present the example of fittings with the GCS technique for two studied events, in order to better demonstrate the orientation of the fitted flux rope. 
For both studied events the faint edge of the WL shock was observed, at least as seen from one viewing point. The GCS fittings were done after carefully inspecting the EUV and other available observations, in order to fit the CME as accurate as possible.

In the presented pairs of images we show on the left hand side the coronagraph image from STEREO A/COR2 and on the right hand side from SOHO/LASCO C2 or C3. 

\begin{figure*}[htb]
    \centering
    \includegraphics[width=0.75\linewidth]{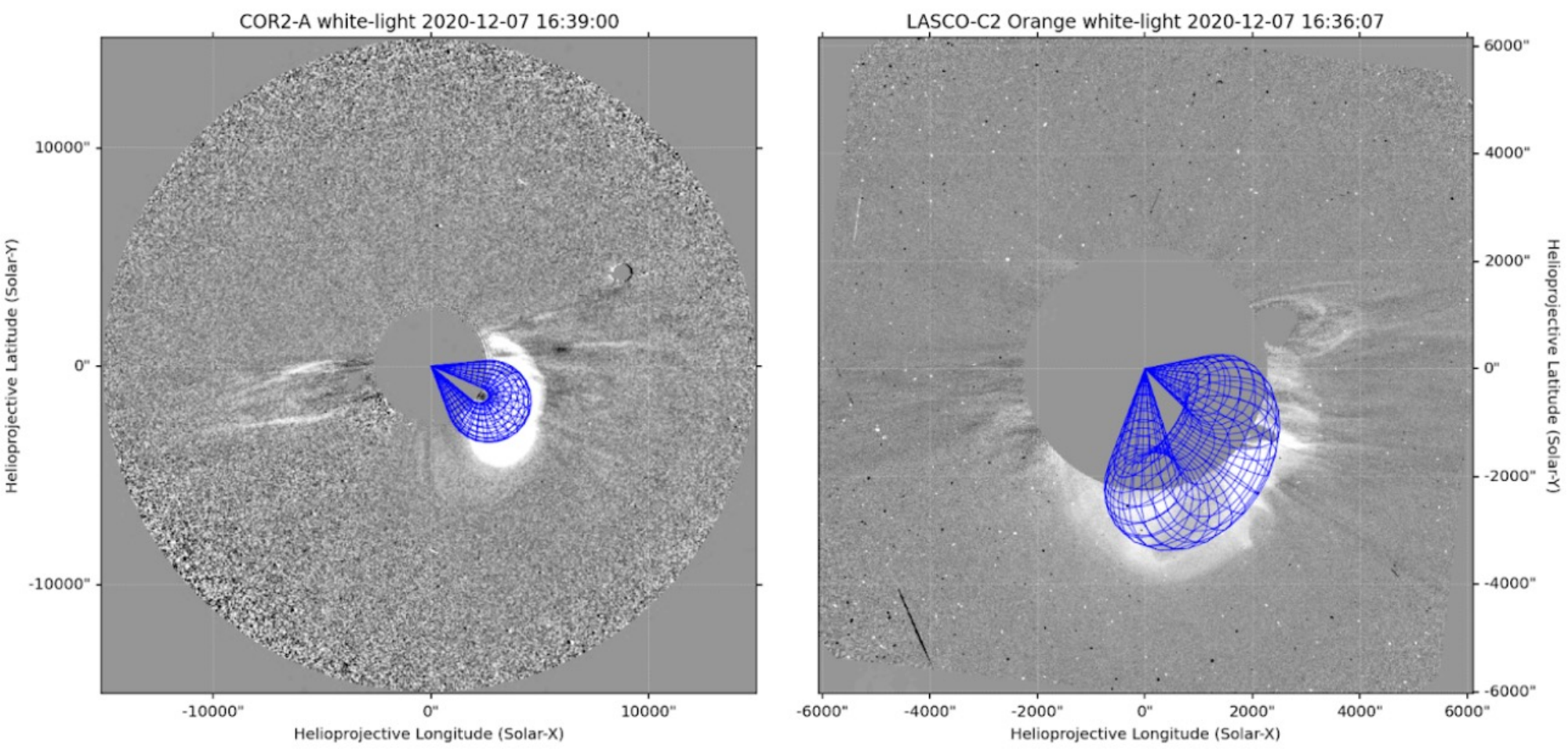}
    \caption{The pair of coronagraph images that we fitted with the GCS technique for GCS2 for Event 1, from STEREO A/COR2 on the left hand side and from SOHO/LASCO C2 on the right hand side. The fitted flux rope is in blue colour.}
    \label{fig:Example_fit_Event2}
\end{figure*}   

\begin{figure*}[htb]
    \centering
    \includegraphics[width=0.784\linewidth]{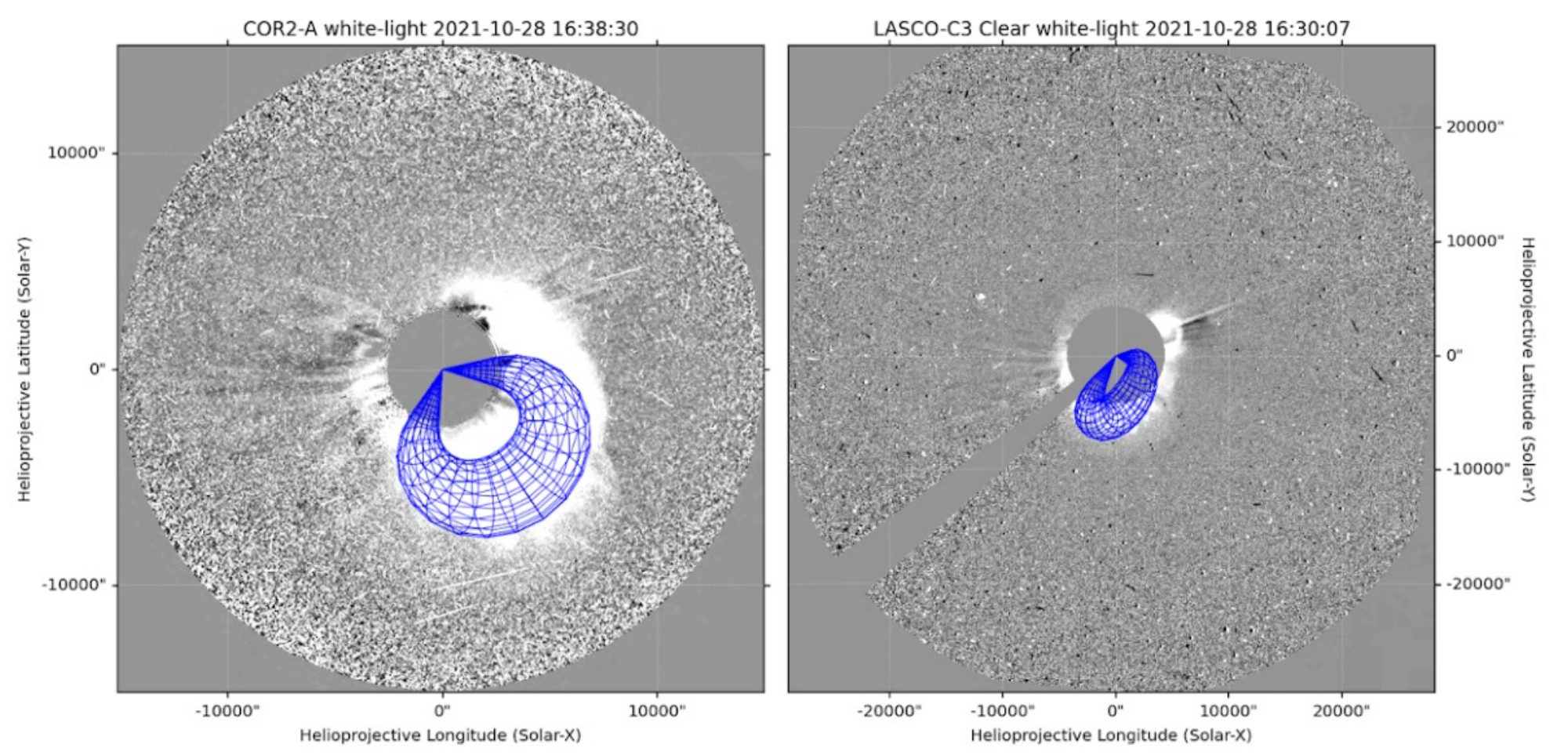}
    \caption{The pair of coronagraph images that we fitted with the GCS technique for GCS3 for Event 2, from STEREO A/COR2 on the left hand side and from SOHO/LASCO C3 on the right hand side. The fitted flux rope is in blue colour.}
    \label{fig:Example_fit_Event1}
\end{figure*} 

\end{appendix}

\end{document}